\documentclass[aps,prd,twocolumn,superscriptaddress,nofootinbib]{revtex4-2}
\usepackage{amsmath}
\usepackage{cases}
\usepackage{amsfonts}
\usepackage{amssymb}
\usepackage{graphicx}
\usepackage{float}
\usepackage{xcolor}
\usepackage{dsfont}
\usepackage{enumitem}
\DeclareUnicodeCharacter{2212}{-}

\newcommand{\K}{{\cal K}}

\begin{document}
\title{On anisotropic two-fluid stellar objects in General Relativity}
\author{Nolene F. Naidu}
\email[]{nolene.naidu@physics.org}
\affiliation{Department of Mathematics and Applied Mathematics, University of Cape Town}
\author{Sante Carloni}
\email[]{sante.carloni@unige.it}
\affiliation{DIME Sez. Metodi e Modelli Matematici, Universit\`{a} di Genova,\\ Via All'Opera Pia 15, 16145 - Genoa, (Italy).}
\author{Peter Dunsby}
\affiliation{Department of Mathematics and Applied Mathematics, University of Cape Town}
\date{\today}

\begin{abstract}
We apply the 1+1+2 covariant semi-tetrad approach to describe a general static and spherically symmetric relativistic stellar object which contains two fluids with anisotropic pressure. The corresponding Tolman-Oppenheimer-Volkoff equations are then obtained in covariant form for the anisotropic case. These equations are used to obtain new exact solutions using direct resolution and reconstruction techniques.  Finally, we show that three of the generating theorems known for the single fluid case can also be used to obtain two-fluid solutions from single fluid ones. 
\end{abstract}
\maketitle
%%%%%%%%%%%%%%%%%%%%%%%%%%%%%%%%%%
\section{Introduction}
%%%%%%%%%%%%%%%%%%%%%%%%%%%%%%%%%%
Studying the internal structure of stars and compact objects in a relativistic context is an important area of relativistic astrophysics. It is, however, notoriously complex, particularly when approached using analytic methods. The fundamental equations that describe the relativistic hydrostatic equilibrium, the Tolman-Oppenheimer-Volkoff (TOV) equations \cite{tolman,oppvolk}, though quasilinear, present many inherent challenges in their analytical resolution. To date, only a few physically relevant exact solutions have been found, some of them almost a century-old \cite{del}.

The challenge is intensified by the fact that realistic compact objects have a structure that comprises different regions with wildly different properties. For example, neutron stars possess an inner and outer core as well as an inner and outer crust; the compositions of which range from standard matter to superfluid neutronium \cite{jurgen}. The differences in composition and conditions between these regions are so marked that it is not advantageous to attempt to describe them using a single equation of state. We suppose, instead, that these compact objects comprise different fluids, each with its own equation of state. In this picture, we expect neutron stars and other relativistic compact objects to present a so-called ``shell'' structure in which every layer is composed of different fluids or a mixture of fluids. Since we will have a different solution of the TOV equations for each of these fluids, it becomes crucial to join these solutions to obtain a complete solution for the entire object. The key tool for this task is the junction conditions, i.e., the conditions necessary to solder together different spacetimes. There are different formulations of the junction conditions, but the most used is the Israel formulation \cite{Israel:1966rt}. These conditions will also be used to match the solutions associated with the interior of the compact object to the exterior, typically represented by the vacuum Schwarzschild \cite{1916skpa.conf.424S}, or the radiating Vaidya \cite{vaidya} solutions. In the following, we will use the term ``complete'' solution for any solution obtained in this way, so that it describes the entire spacetime associated with the interior of a given compact object.

Another important aspect to consider in a realistic model of compact stellar objects is the inclusion of anisotropies, arising from the difference between radial and transverse pressures. There are many possible reasons for the existence of anisotropy in astrophysical objects, such as strong electromagnetic fields and the exotic thermodynamical properties in the state of matter \cite{herrera} (and references therein). Letelier \cite{letelier} also proposed that a mixture of real gas can exhibit anisotropy.

Anisotropy adds extra degrees of freedom that make the search for analytical anisotropic interior solutions for these objects highly non-trivial. Compact objects which admit local anisotropies present modifications to their physical parameters that affect the redshift, critical mass, and the stability of the stellar objects \cite{viaggiu}. Exotic objects, such as boson stars and generalized classes of gravastars, are intrinsically anisotropic in pressure \cite{visser}. In general, for a mixture of two perfect fluids in relative motion, the total energy-momentum tensor in the comoving frame of either fluid is anisotropic in the pressure \cite{letelier}.

There has been steady incremental progress on the analytical study of anisotropic solutions of the TOV equations. Harko and Mak \cite{harko} assume a specific value for the anisotropy parameter and give a class of physically viable, exact solutions. Rago \cite{rago} introduced an anisotropic function to generalize the Tolman V solution to include anisotropy, and a generating function, forming part of the foundation for the generating theorems in \cite{boon1, boon2}. The work by Viaggiu \cite{viaggiu} is a perturbative analysis applied to known ''seed" solutions such as Florides and Tolman IV. In the paper by Das {\it et al}. \cite{das}, the anisotropic extension to Tolman IV is given. Their model is compared to pulsar observational data and satisfies the physical viability conditions given in \cite{del}.

The most substantial advances in this research line, however, have been achieved numerically. There has been relevant progress with simulations and approximations in this regard (see e.g. \cite{numerics1}-\cite{numerics6} and references therein). In this context, analytical studies of the TOV equations are important in addition to numerical studies for several reasons. Exact solutions can be used to explore the full parameter space for a given metric rather than a single set of values. They can also be utilized to test numerical codes, particularly when they involve new approximations schemes. Furthermore, analytical solutions can be tested using observational data to constrain the parameters in these models \cite{das}.

In this paper, we will construct a covariant formulation of the TOV equations describing a multifluid compact object with anisotropies. This work will be an extension of \cite{ncd1}, in which we developed the covariant TOV equations for multifluid isotropic compact objects. We explore direct resolution strategies to obtain two-fluid solutions from known single fluid ones with consideration of the energy conditions to ensure the obtained solutions are physically viable. In order to be considered physically viable, the behaviour of the thermodynamical quantities in these solutions must satisfy the constraints given for at least one set of the parameters. We extend the reconstruction algorithms to include anisotropy and give an example of a solution, constructed by using a combination of known single fluid solutions, the Tolman IV and the interior Schwarzschild solutions, applied to the two-fluid model. In Boonserm {\it et al}. \cite{boon1, boon2} it was shown that one can obtain new single fluid solutions from known (seed) solutions using generating theorems. Here, we give three examples of variations to the generating theorems, and show that they can be applied to the two-fluid anisotropic case.

The cornerstone of this paper is the covariant 1+1+2 approach. This formalism was pioneered by Clarkson and Barrett in 2003 \cite{clarkbar}, where it was applied to the study of perturbations of Schwarzschild black holes. The fact that this approach is ideally suited to the study of spherically symmetric systems and their perturbations led Carloni to apply this framework to study single fluid isotropic \cite{sante1} and anisotropic stellar models \cite{sante2}, leading to physically viable exact solutions in both cases.

The following analysis involves simplifying the structure described above by using one or two fluids with a given equation of state. It is relatively easy to generalise the two fluid equations to the case of the $N$ fluids. While such generalisation would provide the opportunity to examine a wider array of objects, it is well known that neutron stars require just two fluids to be described at a satisfactory degree of accuracy \cite{lang}. The cases with more than two fluids can be explored to potentially give a higher degree of accuracy to models in the future. In order to give a truly complete and realistic description, we would have to include other characteristics such as rotation, phase transitions, charge, flux, interactions, etc. While the two-fluid equations obtained can be easily generalized to include some of these effects, we will not include such factors in this work. These aspects will be the subject of future research.

The outline of this paper is as follows: In Sec. \ref{TFC} the 1+1+2  semi-tetrad equations for the two-fluid solution are presented. Section \ref{CON} deals with the conditions of physical viability for a given solution of the TOV equations. Section \ref{Junction}, instead, deals with the generalisation of Israel's junction conditions to the multifluid case.  Section \ref{KS} contains the expressions for the Bowers-Liang solution, an element of the class of Florides solutions, the interior Schwarzschild solution, and the Tolman IV solution. In Sec. \ref{RS}, we explore some direct resolution strategies applied to the solutions in Sec. \ref{KS}. Sec. \ref{SR} contains solution reconstruction techniques. The generating theorems of \cite{boon1} are discussed and applied to a two-fluid system in Sec. \ref{GT}. A discussion and some concluding remarks can be found in Sec. \ref{CR}. The main equations of the 1+1+2 formalism are presented in Appendix \ref{AA} and the $N$ fluid generalisation of the equations in Appendix \ref{App2}. Appendix \ref{AARS} contains the full solutions for Sec. \ref{RS}, while Appendix \ref{AASR} has the full expressions for solutions obtained in Sec. \ref{SR}.
%%%%%%%%%%%%%%%%%%%%%%%%%%%%%%%%%
\section{The 1+1+2 equations for the two-fluid case}\label{TFC}
%%%%%%%%%%%%%%%%%%%%%%%%%%%%%%%%%
The 1+1+2 equations can be written easily for any number of fluids (see Appendix \ref{App2} for the extension to $N$ fluids). In this section and what follows, we will consider the application to two fluids. In \cite{lang}, this is shown as sufficient to describe systems such as neutron stars, which are one of the primary applications of the TOV equations. 
We begin by defining a time-like threading vector field $u^a$ that is associated to the observer's congruence with $u^a u_a = −1$, and a space-like vector $e^a$ with $e_a e^a = 1$. The $u^a$ and $e^a$ congruences allow one to foliate the spacetime in hypersurfaces whose geometry is defined by the two projection tensors
\begin{equation}
\begin{split}
{h^a}_b &= {g^a}_b + u^a u_b \hspace{0.2cm},\hspace{0.2cm} {h^a}_a = 3, \\
{N_a}^b &= {h_a}^b - e_a e^b = {g_a}^b + u_a u^b - e_a e^b \hspace{0.1cm},\hspace{0.1cm}{N^a}_a = 2,
\end{split}
\end{equation}
Here ${h^a}_b$ represents the metric of the 3-spaces orthogonal to $u^a$, and ${N_a}^b$ represents the metric of the 2-spaces orthogonal to $u^a$ and $e^a$. Any tensorial object may now be split according to the above foliations, as described in \cite{sante1}. The covariant time derivative, orthogonally projected covariant derivative, hat-derivative and $\delta$-derivative are given by
\begin{eqnarray}
{\dot{X}{}^{a..b}}_{c..d} &\equiv& u^e {\nabla}_e {X^{a..b}}_{c..d}, \nonumber \\
D_e {X^{a..b}}_{c..d} &\equiv& {h^a}_f...{h^b}_g {h^p}_c...{h^q}_d {h^r}_e {\nabla}_r {X^{f..g}}_{p..q}, \nonumber \\
{\hat{X}{}_{a..b}}^{c..d} &\equiv& e^f D_f {X_{a..b}}^{c..d}, \nonumber \\
{\delta}_e {X_{a..b}}^{c..d} &\equiv& {N_a}^f...{N_b}^g {N_i}^c...{N_j}^d {N_e}^p D_p {X_{f..g}}^{i..j}, \hspace{0.3cm}
\end{eqnarray}
respectively.

The full set kinematical variables associated to the covariant derivatives of $u^a$ and $e^a$ is given in Appendix \ref{AA}.

The energy-momentum tensor can be decomposed as
\begin{eqnarray}\label{Tab}
T_{ab} = &&\mu u_a u_b + (p + \Pi) e_a e_b + \left(p - \frac{1}{2} \Pi \right) N_{ab} \nonumber\\
&& + 2 Q e_{(a} u_{b)} + 2 Q_{(a} u_{b)} + 2 \Pi_{(a} e_{b)} + \Pi_{ab},
\end{eqnarray}
where the symmetrisation over the indices of a tensor is represented as $T_{(ab)} = \frac{1}{2} (T_{ab} + T_{ba})$. The matter variables are given by
\begin{eqnarray}
\mu &=& T_{ab} u^a u^b, \nonumber \\
p &=& \frac{1}{3} T_{ab} \left(e^a e^b + N^{ab}\right), \nonumber \\
\Pi &=& \frac{1}{3} T_{ab} \left(2 e^a e^b - N^{ab}\right), \nonumber \\
Q &=& - T_{ab} e^a u^b, \nonumber \\
Q_a &=& -T_{cd} {N^c}_a u^d, \nonumber \\
\Pi_a &=& T_{cd} {N^c}_a e^d, \nonumber \\
\Pi_{ab} &=& T_{\{ab\}},
\end{eqnarray}
where $\mu$ is the density, $p$ is the pressure, $Q$ and $Q_a$ represent the scalar and vector parts of the heat flux, and $\Pi$ and $\Pi_a$ represent the the scalar and vector components of the anisotropic pressure, respectively. The curly brackets $\{ \}$ denote the Projected Symmetric Trace-Free part of a tensor with respect to $e^a$.

The radial and tangential pressures which appear as sources in the gravitational field equations are given by
\begin{subequations}\label{prpt}
\begin{align}
p_r &= p + \Pi, \\
p_{\perp} &= p - \frac{1}{2}\Pi.
\end{align}
\end{subequations}

Spacetimes which have a unique preferred spatial direction at each point and exhibit local rotational symmetry around such direction are called ``LRS spacetimes'' and can be treated easily within the 1+1+2 formalism. A specific class of non-rotating, non-twisting LRS spacetimes called LRS II spacetimes, has the additional property that all the 1+1+2 vectors and tensors and the quantities  $\Omega, \xi, \mathcal{H}$ vanish identically. If one further chooses to consider static, spherically symmetric spacetimes, then $\Theta, \Sigma$ and $Q$ are zero, as well as the dot derivatives of scalar quantities. In the following, we will deal specifically with static and spherically symmetric LRSII spacetimes.

For these spacetimes, the relevant kinematical variables are 
\begin{subequations} 
\begin{align} 
\mathcal{A} &= e_a \dot{u}^a,\\
\phi &= {\delta}_a e^a,   \\
\mathcal{E} &= {C}_{acbd} u^c u^d e^a e^b,
\end{align}
\end{subequations}
where $C_{acbd}$ is the Weyl tensor, and the system (\ref{hatsystem1}) for two fluids with zero total flux, reduces to
\begin{subequations} \label{hatsystem2}
\begin{align} 
\hat{\phi}=&-\frac{1}{2} {\phi}^2 - \frac{2}{3} ({\mu}_1 + {\mu}_2) - \mathcal{E} - \frac{1}{2} ({\Pi}_1  + {\Pi}_2),  \\
\hat{\mathcal{E}}=&  \frac{1}{3} (\hat{\mu}_1 - \hat{\mu}_2) - \frac{1}{2} ( \hat{\Pi}_1 + \hat{\Pi}_2)\nonumber\\
&-\frac{3}{2} {\phi} \left(\mathcal{E} + \frac{1}{2} {\Pi}_1 + \frac{1}{2} {\Pi}_2 \right),  \\ 
-\mathcal{A} \phi &+ \frac{1}{3}({\mu}_1 + 3 p_1) + \frac{1}{3}({\mu}_2 + 3 p_2)\nonumber \\ 
&- \mathcal{E} + \frac{1}{2} {\Pi}_1 + \frac{1}{2} {\Pi}_2 = 0,\\
\hat{\mathcal{A}}=&- \mathcal{A} \left( \mathcal{A} + \phi \right) + \frac{1}{2}({\mu}_1 + 3p_1) + \frac{1}{2}({\mu}_2 + 3p_2), \\
\hat{p}_1 + \hat{\Pi}_1 \label{P1}=&-{\Pi}_1 \left(\frac{3}{2} \phi + \mathcal{A} \right) - \mathcal{A} \left( {\mu}_1 + p_1 \right) + j_e^{(1,2)}, \\
 \hat{p}_2 + \hat{\Pi}_2 =&-{\Pi}_2 \left(\frac{3}{2} \phi + \mathcal{A} \right) - \mathcal{A} \left( {\mu}_2 + p_2 \right) - j_e^{(1,2)}\label{P2} ,\\
K=&\frac{1}{3} ({\mu}_1 + {\mu}_2) - \mathcal{E} - \frac{1}{2} ({\Pi}_1 + {\Pi}_2) + \frac{1}{4} {\phi}^2 ,\\
\hat{Q}_1=&-Q_1 \left(\phi + 2 \mathcal{A} \right) + j_u^{(1,2)},  \label{Q1}\\
Q_2 =& - Q_1, \label{Qeq} 
\end{align}
\end{subequations}
where $ j_u^{(1,2)}$ and $ j_e^{(1,2)}$ represent the interaction terms between fluid 1 and 2.
Next, we introduce a useful parameter, $\rho$, such that $\hat{X} = \phi X_{,\rho}$. In this way, the equation for the Gauss' curvature $K$ \eqref{Keq} can be solved to give
\begin{equation}\label{Krho}
K = K_0^{-1} e^{- \rho}.
\end{equation}
Assuming $\phi\neq 0$\footnote{This assumption is might appear arbitrary. However, it can be shown that $\phi=0$ can be related to the presence of horizons in spacetime. As there will be no horizons in the spacetimes we will consider we can make this assumption without loss of generality.} and defining the variables
\begin{subequations}\label{newvars}
\begin{align}
X &= \frac{{\phi}_{, \rho}}{\phi}, & Y &= \frac{\mathcal{A}}{\phi}, & \mathcal{K} &= \frac{K}{{\phi}^2} ,\\
E &= \frac{\mathcal{E}}{{\phi}^2}, & \mathbb{M}_1 &= \frac{{\mu}_1}{{\phi}^2}, &
\mathbb{M}_2 &= \frac{{\mu}_2}{{\phi}^2}, \\
P_1 &= \frac{p_1}{{\phi}^2}, & P_2 &= \frac{p_2}{{\phi}^2}, &
\mathbb{P}_1 &= \frac{{\Pi}_1}{{\phi}^2} ,\\
\mathbb{P}_2 &= \frac{{\Pi}_2}{{\phi}^2} ,& \mathbb{Q}_1 &= \frac{Q_1}{{\phi}^2}, & 
\mathbb{Q}_2 &= \frac{Q_2}{{\phi}^2} ,\\ 
\mathbb{J}_u &= \frac{j_u^{(1,2)}}{{\phi}^3}, & \mathbb{J}_e &= \frac{j_e^{(1,2)}}{{\phi}^3},
\end{align}
\end{subequations} 
we can recast the system \eqref{hatsystem2} as
\begin{subequations}
\begin{align}
Y_{, \rho} =& - Y ( X + Y + 1) + \frac{1}{2}(\mathbb{M}_1 + \mathbb{M}_2) \nonumber \\ &+ \frac{3}{2}(P_1 + P_2), \label{DE1} \\
\mathcal{K}_{, \rho} =& -\mathcal{K}(1 + 2 X), \label{DE2} \\
{P}_{1, \rho} + \mathbb{P}_{1, \rho} = &- Y (\mathbb{M}_1 + P_1) \nonumber \\
&- \mathbb{P}_1 \left( 2X + Y + \frac{3}{2}\right) - 2 X P_1 + \mathbb{J}_e, \label{fl1_cons}
\end{align}
\begin{align}
{P}_{2, \rho} + \mathbb{P}_{2, \rho} = &- Y (\mathbb{M}_2 + P_2) \nonumber \\
&- \mathbb{P}_2 \left( 2X + Y + \frac{3}{2}\right) - 2 X P_2 - \mathbb{J}_e, \label{fl2_cons}\\
\mathbb{Q}_{1, \rho}=& -\mathbb{Q}_1 ( 1 + 2 X + 2 Y) + \mathbb{J}_u,
\end{align}
\end{subequations}
with the following constraints
\begin{eqnarray}\label{constraints}
2(\mathbb{M}_1 + \mathbb{M}_2) + 2 (P_1 + P_2) + 2 (\mathbb{P}_1 &&+ \mathbb{P}_2)+ 2 X \nonumber \\ 
&& - 2 Y + 1 = 0,  \label{c1}\\
1 - 4 \mathcal{K} - 4 (P_1 + P_2) + 4 Y - 4 (\mathbb{P}_1 &&+ \mathbb{P}_2) = 0, \label{c2}\\
2(\mathbb{M}_1 + \mathbb{M}_2)+ 6 (P_1 + P_2) + 3 (\mathbb{P}_1 &&+ \mathbb{P}_2) \nonumber \\
&&- 6 Y - 6 E = 0. \label{c3}
\end{eqnarray}
Rearranging \eqref{c3}, \eqref{c1}, and writing \eqref{Qeq} in terms of $\rho$, we can write
\begin{eqnarray}\label{constraints2}
E &=& \frac{1}{3} \left( \mathbb{M}_1 + \mathbb{M}_2 \right) + P_1 + P_2 + \frac{1}{2} \left(\mathbb{P}_1 + \mathbb{P}_2 \right) - Y, \\
X &=&  -\frac{1}{2} - (\mathbb{M}_1 + \mathbb{M}_2) - (\mathbb{P}_1 + \mathbb{P}_2) \nonumber \\
&&-  (P_1 + P_2) + Y, \label{constrX}\\
\mathbb{Q}_{1}&=&-\mathbb{Q}_{2}, \label{constrQ}
\end{eqnarray}
where differentiation with respect to $\rho$ is indicated by a comma and the subscript $\rho$. 
The equation for the total pressure is given by
\begin{equation}
\begin{split}
{P}_{tot, \rho}  + \mathbb{P}_{tot, \rho} &+  P_{tot}(Y+2X) + Y \mathbb{M}_{tot}\\
&+ \mathbb{P}_{tot}  \left( 2X + Y + \frac{3}{2}\right) =0  , \label{DE3}
\end{split}
\end{equation}\\
where, from now on the subscript ``$tot$'' of a given quantity represents the total or sum of this quantity, for fluids 1 and 2. So, for example $P_{tot}=P_{1}+P_2$. We can write the covariant equivalent of the TOV equations by using the constraints to eliminate all the metric-related variables except $\mathcal{K}$, resulting in
\begin{widetext}
\begin{eqnarray}\label{tov2} 
{P}_{1, \rho} + \mathbb{P}_{1, \rho}   &=& \mathbb{J}_e- {P_1}^2  - {\mathbb{P}_1}^2 + P_1 \left[\mathbb{M}_1 - 2\mathbb{P}_1 - 3 \mathcal{K} + \frac{7}{4} \right] + \mathbb{P}_1 \left(\mathbb{M}_1 - 3\mathcal{K} + \frac{1}{4}\right) + \mathbb{M}_1 \left(\frac{1}{4} -\mathcal{K}\right) \nonumber \\
&&  - P_1 (P_2 +  \mathbb{P}_2 -  2\mathbb{M}_2) - \mathbb{P}_1 (P_2+ \mathbb{P}_2 - 2\mathbb{M}_2)-\mathbb{M}_1 \left(P_2 +  \mathbb{P}_2\right), \label{p1}\\
{P}_{2, \rho} + \mathbb{P}_{2, \rho}  &=& -\mathbb{J}_e - {P_2}^2  - {\mathbb{P}_2}^2 + P_2 \left[\mathbb{M}_2 - 2\mathbb{P}_2
- 3 \mathcal{K} + \frac{7}{4} \right] + \mathbb{P}_2 \left(\mathbb{M}_2 - 3\mathcal{K} + \frac{1}{4}\right) + \mathbb{M}_2 \left(\frac{1}{4} -\mathcal{K}\right) \nonumber \\
&&  - P_2 (P_1 +  \mathbb{P}_1 - 2\mathbb{M}_1) - \mathbb{P}_2 (P_1 +  \mathbb{P}_1 - 2\mathbb{M}_1)-\mathbb{M}_2 \left(P_1 +  \mathbb{P}_1\right), \label{p2}\\
\mathcal{K}_{, \rho} &=&  2 \mathcal{K} \left( \frac{1}{4}-\mathcal{K} + \mathbb{M}_1 + \mathbb{M}_2 \right), \label{tovk} \\
\mathbb{Q}_{1, \rho}&=& \mathbb{Q}_1 \left[ 2\mathcal{K} - 2 (\mathbb{M}_1 + \mathbb{M}_2)-\frac{3}{2} \right]- \mathbb{J}_u,\\
{P}_{tot, \rho} + \mathbb{P}_{tot, \rho} &=& - {P_{tot}}^2  - {\mathbb{P}_{tot}}^2 + P_{tot} \left[\mathbb{M}_{tot} - 2 \mathbb{P}_{tot} - 3 \mathcal{K} + \frac{7}{4} \right] + \mathbb{P}_{tot}  \left[\mathbb{M}_{tot}  - 3 \mathcal{K} + \frac{1}{4} \right] + \mathbb{M}_{tot}\left( \frac{1}{4} - \mathcal{K} \right). \label{EqPtot}
\end{eqnarray}
\end{widetext}

The $N$ fluid generalization of the above equations can be found in Appendix \ref{App2}. The above equations require an equation of state, relating the pressure and density, in order to be complete. However, in the following we will not assume a form for the equation of state unless it is absolutely necessary. In this way we will be able to define several methods to solve the covariant TOV equations (\ref{tov2}-\ref{EqPtot}).

In the following sections, we will present solutions of the above equations. In the interest of aiding the physical interpretation of these solutions, we present the following useful relations. For a generic metric tensor of the form 
\begin{align}\label{metricgen}
ds^2 =& -k_1(x,t)  dt^2 + k_2(x,t) dx^2 \nonumber \\
&+ k_3(x,t) \left[dy^2 + k_4(y) dz^2\right],
\end{align}
with spacetime geometry represented by
\begin{align}
k_4(y) &= \left\{ \begin{array}{ll}
\sin{y},&\mbox{closed geometry}\\
y, &\mbox{flat geometry}\\
\sinh{y},&\mbox{open geometry}
\end{array}
\right.,
\end{align}
we can write \cite{betschart}
\begin{equation}
\begin{split}\label{AphiCoord}
\phi=\frac{{k}_{3,x}}{k_2 k_3},\quad
\mathcal{A}=\frac{{k}_{1,x}}{2 k_2 k_1}.
\end{split}
\end{equation}
As we will mainly use the parameter $\rho$ in our calculations, it is useful to give the same formulae for the case in which the metric is written as
\begin{equation}\label{metrickRho}
ds^2 = -k_1(\rho) dt^2 + k_2(\rho) d\rho^2 + k_3(\rho) d\Omega^2,
\end{equation}
where 
\begin{align}
k_3(\rho) &= K_0 e^{\rho},\\
d\Omega^2 &= d\theta^2 + \sin^2 \theta d\phi^2.
\end{align}
In terms of $\rho$, we then have
\begin{equation} \label{vars}
\begin{split}
\phi&= \frac{1}{\sqrt{k_2}},\quad
\mathcal{A}= \frac{k_{1,\rho}}{2 k_1\sqrt{k_2}},\\
\mathcal{K} &=  \frac{k_2}{K_0  e^{\rho}},\quad
Y = \frac{k_{1,\rho} }{2 k_1}, 
\end{split}
\end{equation}
where $K_0$ is a suitable constant.  With the above relation, it is not difficult to express the newly obtained solutions in terms of the area radius, $r$. The relation between $r$ and $\rho$  is 
\begin{align}\label{rrho}
\rho &= 2 \ln \left(\frac{r}{r_0}\right),\\
r &=\sqrt{K_0} e^{\rho/2},
\end{align}
where $r_0$ and $K_0$ are constants related by $K_0=r_0^2$. 

The conversion from the metric coefficients in the $\rho$ coordinate to the $r$ coordinate can be achieved by noting that $k_1$ and $k_3$ are scalars with respect to a change of the radial parameter and that 
\begin{equation}
k_2 (\rho)\rightarrow \frac{r^2}{4}k_2 (r).
\end{equation}
Therefore, in terms of the area radius, we have
\begin{equation}\label{metrick_r}
ds^2 = -k_1(r) dt^2 + k_2(r) dr^2 + r^2 d\Omega^2,
\end{equation}
and 
\begin{equation} \label{vars_r}
\begin{split}
\phi&=\frac{2}{r\sqrt{k_2}},\quad
\mathcal{A}= \frac{k_{1,r}}{2 k_1\sqrt{k_2}},\\
\mathcal{K} &=  \frac{k_2}{4},\quad
Y = \frac{k_{1,r} }{4 k_1}. 
\end{split}
\end{equation}
These formulae will be useful to write the results of the TOV equations in a form closer to the usual one in literature.
%%%%%%%%%%%%%%%%%%%%%%%%%%%%%%%%%%%%%
\section{Conditions for physical viability}\label{CON}
%%%%%%%%%%%%%%%%%%%%%%%%%%%%%%%%%%%%%
Though many solutions to the TOV equations exist, several of these are not physical. In this section, we provide a set of conditions that a solution must satisfy in order to be physically relevant \cite{del,Carloni:2014rba,sante1,sante2}. 

We require:
\begin{enumerate}[label=(\roman*)]
\item the sources of the Einstein equations are positive definite
\begin{equation}\label{pcond}
\mu \geq 0,\quad p + \Pi \geq 0, \quad  p -\frac{1}{2} \Pi \geq 0,
\end{equation}

\item both fluids (individually and combined) satisfy the weak energy  condition
\begin{equation}\label{energycond}
\mu \geq 0, \quad \mu  + p + \Pi \geq 0, \quad \mu + p -\frac{1}{2} \Pi\geq 0,
\end{equation}

\item each fluid satisfies the conditions 
\begin{equation}\label{stabcond}
\mu^{\prime} < 0, \quad p^{\prime}+\Pi^{\prime} < 0,  
\end{equation}
where the prime represents the derivative with respect to the area radius. These conditions are necessary (but not sufficient) for the stability of the solution. If $p$ is a decreasing, and $\mu$ is increasing, we have that $\frac{dp}{d\mu}$ is negative. The requirement of a negative gradient for the pressure and density is equivalent to the requirement that the speed of sound is real.

\item for both fluids causality, is preserved and therefore the speed of sound must never be higher than the speed of light:
\begin{equation}\label{soundcond}
\begin{split}
0< c_{s,r}^2&+c_{s,\perp}^2\leq 1,
\end{split}
\end{equation}
where 
\begin{equation}\label{soundconds}
\begin{split}
c_{s,r}^2&= \frac{\partial p_r}{\partial \mu},\\
c_{s,\perp}^2&=\frac{\partial p_\perp}{\partial \mu},
\end{split}
\end{equation}
\item the 1+1+2 potentials and therefore, the metric coefficients and some of their first derivatives are finite and positive valued at the center of the matter distribution.
\end{enumerate}

Notice that these conditions do not limit the sign of the anisotropy $\Pi$. In fact, combining \eqref{energycond} we obtain
\begin{equation}\label{ecnew}
-p\leq \Pi \leq 2p,  
\end{equation}
which shows that the anisotropic pressure can be negative without violating the weak energy condition. 

In the case of a complete solution, we consider the additional requirement that the total radial pressure is decreasing. Instead, the total anisotropic pressure can have any behaviour compatible with the conditions above. 

%%%%%%%%%%%%%%%%%%%%%%%%%%%%%%%%%%%%%%%%%%%%%
\section{Junction conditions}\label{Junction} 
%%%%%%%%%%%%%%%%%%%%%%%%%%%%%%%%%%%%%%%%%%%%%
An important aspect in the search for solutions is the problem of the junction between the interior solution and the exterior vacuum spacetime, which is characterised by the Schwarzschild metric. The procedure to determine the junction conditions is similar to the single fluid case treated in \cite{sante1,sante2}. In particular, Israel's junction conditions~\cite{Israel:1966rt,Barrabes:1991ng} are equivalent to
\begin{equation}\label{Israel2}
[\K]=0, \qquad [Y]=0\,.
\end{equation}
Using the constraint Eq.~\eqref{c2} in \eqref{Israel2}, we obtain
\begin{equation}\label{JunCov}
 \left[P_1+\mathbb{P}_1+P_2+\mathbb{P}_2\right]=0\,.
\end{equation}
This implies that in order to provide a smooth junction with the Schwarzschild metric, the sum of the radial pressure for both fluids must be zero at the junction. As the radial pressures are positive definite in a realistic solution, this result implies that the radial pressure of {\it both} fluids must be zero at the junction. A possible occurrence in a two-fluid system is that the radial pressure of one fluid goes to zero before the other. In this case, we will consider our solution as valid only when both pressures are positive, regardless of what happens afterwards. Such solutions will be referred to as representing a {\em shell} in a compact stellar object\footnote{It should be noted that the vanishing of the radial pressure on the boundary is more of a desirable feature than a physical necessity. In fact, there is no requirement for a ``hard boundary'' in a stellar object, provided that the total mass remains finite. Nevertheless, in the following we will always require that our solutions present a hard boundary.}. 

%%%%%%%%%%%%%%%%%%%%%%%%%%%%%%%%%%%%%%%%%%%%%%%%%%%%%%%%%%%%%%%%%%%%%%%%%%%%%%
\section{Some known single fluid solutions of the TOV equations}\label{KS}
%%%%%%%%%%%%%%%%%%%%%%%%%%%%%%%%%%%%%%%%%%%%%%%%%%%%%%%%%%%%%%%%%%%%%%%%%%%%%%
In the sections that follow, we will consider several direct and indirect resolution methods of the TOV equations \eqref{tov2}. These methods rely on known single fluid solutions. In this section, we introduce four solutions in a form compatible with the formalism we employ: two anisotropic solutions, the Bowers-Liang \cite{bowers} and an element in the class of Florides \cite{florides} solutions, as well as two isotropic solutions: the interior Schwarzschild solution \cite{1916skpa.conf.424S}, and the Tolman IV solution \cite{tolman}. 
%%%%%%%%%%%%%%%%%%%%%%%%%%%%%%%%%%%%%%%%%%%%%%%%%%%%%%%%%%%%%%%%%%%%%%%%%%%%%%%%%%%%%%%%%%%%%%
\subsection{Bowers-Liang solution}
%%%%%%%%%%%%%%%%%%%%%%%%%%%%%%%%%%%%%%%%%%%%%%%%%%%%%%%%%%%%%%%%%%%%%%%%%%%%%%%%%%%%%%%%%%%%%%
The Bowers-Liang solution, discovered in 1974 \cite{bowers} is the extension of the interior Schwarzschild solution \cite{1916skpa.conf.424S} to include anisotropy. In this subsection, we present the single fluid Bowers-Liang solution.

For a metric written in the form \eqref{metrick_r}, the Bowers-Liang solution is given as
\begin{equation} \label{blsoln}
\begin{split}
k_1 &= A_0 \left[3^{\frac{h}{2}+1} \mathfrak{c}_1 + \left(3-\mu_0 r^2\right)^{\frac{h}{2}}\right]^{\frac{2}{h}}, \\
k_2 &= \frac{3}{3 - \mu_0 r^2},\\
k_3 &= r^2,
\end{split}
\end{equation}
where $A_0$, ${\mathfrak c}_{1}$, $h$ and $\mu_0$ are constants. The metric represented by \eqref{blsoln} corresponds, via the Einstein equations, to the following expressions for the radial pressure, tangential pressure and energy density respectively
\begin{subequations}\label{blmv}
\begin{align}
{p_r}^{BL}(r) &= -\frac{\mu_0 \left(3^{-\frac{h}{2}} \mathfrak{z}_0^{\frac{h}{2}}+\mathfrak{c}_1\right)}{3^{-\frac{h}{2}} \mathfrak{z}_0^{\frac{h}{2}}+3 \mathfrak{c}_1}, \label{blpr}\\
{p_{\perp}}^{BL}(r) &= -\frac{\mu_0}{\mathfrak{z}_0 \left(3^{\frac{h}{2}+1} \mathfrak{c}_1+\mathfrak{z}_0^{\frac{h}{2}}\right)^2} \left(\mathfrak{z}_0^{h+1} \right. \nonumber\\
& \left. +\mathfrak{c}_1 \left\{3^{h+1} \mathfrak{c}_1 \mathfrak{z}_0+3^{1-\frac{h}{2}} \mathfrak{z}_0^{\frac{h}{2}} \left[12-(h+3) \mu_0 r^2\right]\right\}\right), \label{blpt}\\
\mu_{BL} (r) &= \mu_0,
\end{align}
\end{subequations}
where
\begin{equation}\label{z0r}
\mathfrak{z}_0 = 3-r^2 \mu_0 .
\end{equation}
In terms of the newly defined variables \eqref{newvars} and the parameter $\rho$, we have 
\begin{equation} \label{phibl}
\phi_{BL} = 2 \sqrt{\frac{z}{3 e^{\rho} K_0}},
\end{equation}
and the system \eqref{blmv} corresponds to
\begin{subequations}\label{blnmv}
\begin{align}
\mathcal{K}_{BL}(\rho) =& \frac{3}{4 z_0}, \label{kbl}\\
P_{BL}(\rho) =& \frac{e^{\rho} K_0 \mu_0}{4 z_0^2 \left(3^{-\frac{h}{2}} z_0^{\frac{h}{2}}+3 \mathfrak{c}_1\right)^2} \left\{ -3^{1-h} z_0^{h+1}-9 \mathfrak{c}_1^2 z_0 \right. \nonumber\\
& \left. + 2 \mathfrak{c}_1 3^{-\frac{h}{2}} z_0^{\frac{h}{2}} \left[(h+5) e^{\rho} K_0 \mu_0-18\right]\right\}, \\
\mathbb{P}_{BL}(\rho) =& -\frac{3^{-\frac{h}{2}} (h-1) e^{2 \rho} K_0^2 \mu_0^2 \mathfrak{c}_1 z_0^{\frac{h}{2}-2}}{2 \left(3^{-\frac{h}{2}} z_0^{\frac{h}{2}}+3 \mathfrak{c}_1\right)^2} , \label{ppbl}\\
\mathbb{M}_{BL}(\rho) =& e^{\rho} K_0 \mu_0 \left(\frac{3}{4 z_0}\right), \\
Y_{BL}(\rho) =& -\frac{e^{\rho} K_0 \mu_0 z_0^{\frac{h}{2}-1}}{2 \left(3^{\frac{h}{2}+1} \mathfrak{c}_1+z_0^{\frac{h}{2}}\right)},
\end{align}
\end{subequations}
where
\begin{equation}\label{z0}
z = 3-e^{\rho} K_0 \mu_0 .
\end{equation}
%%%%%%%%%%%%%%%%%%%%%%%%%%%%%%%%%%%%%%%%%%%%%%%%%%%%%%%%%%%%%%%%%%%%%%%%%
\subsection{A Florides-class solution}\label{Flor}
%%%%%%%%%%%%%%%%%%%%%%%%%%%%%%%%%%%%%%%%%%%%%%%%%%%%%%%%%%%%%%%%%%%%%%%%%
The Florides solution, discovered in 1974, describes a static, spherically symmetric distribution of dust, that has vanishing radial pressure but nonzero tangential pressure \cite{florides}. It represents the simplest element in a class of solutions of the TOV equations for which
\begin{equation}\label{fpp}
\mathbb{P} = \frac{1}{6}\mathbb{M}\left(1- 4\mathcal{K}\right),
\end{equation}
holds \cite{sante2}. In this subsection, we present an example that belongs to the class of single fluid Florides solutions characterized by a constant density.

This solution can be written in the form (\ref{metrick_r}) setting
\begin{equation} \label{fsoln}
\begin{split}
k_1 &=\frac{B_0 \left(e^{A_0}+\frac{r^2}{r_0^2}\right)^2}{\sqrt{3-\mu_0 r^2}}, \\
k_2 &= \frac{3}{3-\mu_0 r^2},\\
k_3 &= r^2,
\end{split}
\end{equation}
where $A_0$, $B_0$ and $\mu_0$ are constants. As mentioned previously, $r_0$ is a constant. The metric represented by \eqref{fsoln} corresponds, via the Einstein equations, to the following expressions for the radial pressure, tangential pressure and energy density respectively
\begin{subequations}\label{fmv}
\begin{align}
{p_r}^{F}(r) &=  -\frac{4 \left(\mu_0 r^2-3\right)}{3 \left(e^{A_0} r_0^2+r^2\right)}, \label{fpr}\\
{p_{\perp}}^{F}(r) &= \frac{\mu_0^2 r^2}{4\left(3- \mu_0 r^2\right)} - {p_r}^{F}, \label{fpt}\\
\mu_{F} (r) &= \mu_0.
\end{align}
\end{subequations}
In terms of the newly defined variables \eqref{newvars} and the parameter $\rho$, we have 
\begin{equation}
\phi_{F} = -2 \sqrt{\frac{3-e^{\rho} K_0 \mu_0}{3 e^{\rho} K_0}} \label{phif},
\end{equation}
and the system \eqref{fmv} corresponds to
\begin{subequations}\label{fnmv}
\begin{align}
\mathcal{K}_{F}(\rho) =& \frac{3}{4 z}, \label{K_F}\\
P_{F}(\rho) =& \frac{e^{2 \rho} K_0^2 \mu_0^2}{8 z^2}+\frac{1}{1+e^{-\rho+\mathfrak{c}_1}}, \\
\mathbb{P}_{F}(\rho) =&  -\frac{e^{2 \rho} K_0^2 \mu_0^2}{8 z^2}, \\
\mathbb{M}_{F}(\rho) =& e^{\rho} K_0 \mu_0 \left(\frac{3}{4 z}\right), \\
Y_{F}(\rho) =& e^{\rho} \left(\frac{K_0 \mu_0}{4 z}+\frac{1}{e^{\rho}+e^{\mathfrak{c}_1}}\right),
\end{align}
\end{subequations}
where $\mathfrak{c}_1$ is a constant and we have used \eqref{z0}. 
%%%%%%%%%%%%%%%%%%%%%%%%%%%%%%%%%%%%%%%%%%%%%%%%%%%%%%%%%%%%%%%%%%%%%%%%%%%%%%%%%%%%%%%%%%%%%%
\subsection{The interior Schwarzschild solution}\label{intsch}
%%%%%%%%%%%%%%%%%%%%%%%%%%%%%%%%%%%%%%%%%%%%%%%%%%%%%%%%%%%%%%%%%%%%%%%%%%%%%%%%%%%%%%%%%%%%%%
The interior Schwarzschild solution \cite{1916skpa.conf.424S}, the first solution for the interior of a static spherically symmetric relativistic object can be written in the form \eqref{metrick_r} by choosing
\begin{equation} \label{cdsoln}
\begin{split}
k_1 &= a_0 \left(c_{1}+\sqrt{3-r^2 \mu_1}\right)^2, \\
k_2 &= \frac{3}{3-r^2 \mu_1},\\
k_3 &= r^2,
\end{split}
\end{equation}
where $a_0$, $c_{1}$ and $\mu_1$ are constants. For the metric described by \eqref{cdsoln}, we have the following expressions for the pressure and energy density, respectively
\begin{subequations}\label{ismv}
\begin{align}
p_{IS}(r) &= \frac{\mu_1 \left(c_1 \sqrt{3-\mu_1 r^2}-3 \mu_1 r^2+9\right)}{3 \mu_1 r^2-9-3 c_1 \sqrt{3-\mu_1 r^2}} , \label{ispr}\\
\mu_{IS} (r) &= \mu_1.
\end{align}
\end{subequations}
In terms of the parameter $\rho$, we have
\begin{equation}\label{phiis}
\phi_{IS} = \frac{2 e^{-\frac{\rho }{2}} \sqrt{3-K_0 \mu_1 e^{\rho }}}{\sqrt{3} \sqrt{K_0}},
\end{equation}
and (\ref{ismv}) corresponds to
\begin{subequations}\label{isnmv}
\begin{align}
\mathcal{K}_{IS}(\rho) =& \frac{3}{12-4 K_0 \mu_1 e^{\rho }},\\
P_{IS}(\rho) =& \frac{K_0 \mu_1 e^{\rho } \left(c_1^2+2 c_1 \sqrt{3-K_0 \mu_1 e^{\rho }}+3 K_0 \mu_1 e^{\rho }-9\right)}{4 \left(K_0 \mu_1 e^{\rho }-3\right) \left(c_1^2+K_0 \mu_1 e^{\rho }-3\right)}, \\
\mathbb{M}_{IS}(\rho) =& \frac{3 \mu_1 e^{\rho }}{12 K_0-4 \mu_1 e^{\rho }}, \\
Y_{IS}(\rho) =& \frac{K_0 \mu_1 e^{\rho }}{-2 c_1 \sqrt{3-K_0 \mu_1 e^{\rho }}+2 K_0 \mu_1 e^{\rho }-6}.
\end{align}
\end{subequations}

%%%%%%%%%%%%%%%%%%%%%%%%%%%%%%%%%%%%%%%%%%%%%%%%%%%%%%%%%%%%%%%%%%%%%%%%%%%%%%%%%%%%%%%%%%%%%%
\subsection{The Tolman IV solution}\label{tolman4}
%%%%%%%%%%%%%%%%%%%%%%%%%%%%%%%%%%%%%%%%%%%%%%%%%%%%%%%%%%%%%%%%%%%%%%%%%%%%%%%%%%%%%%%%%%%%%%
The Tolman IV solution presented in 1939 \cite{tolman}, is a well-known solution with no singularity at $r=0$. The Tolman IV solution corresponds to the choice
\begin{equation} \label{tol4}
\begin{split}
k_1 &= B^2 \left(1 + \frac{r^2}{A^2}\right), \\ 
k_2 &= \frac{R^2(A^2 + 2 r^2)}{\left(R^2 - r^2\right) \left(A^2 + r^2 \right)}, \\ 
k_3 &= r^2 ,
\end{split}
\end{equation}
in the metric (\ref{metrick_r}), where $A$, $B$ and $R$ are constants. For the metric described by \eqref{tol4}, the pressure and energy density are given by
\begin{subequations}\label{tol4mv}
\begin{align}
p_T(r) &= \frac{R^2-A^2-3 r^2}{R^2 \left(A^2+2 r^2\right)}, \label{tol4p}\\
\mu_T (r) &= \frac{R^2 \left(3 A^2+2 r^2\right)+7 A^2 r^2+3 A^4+6 r^4}{R^2 \left(A^2+2 r^2\right)^2}.
\end{align}
\end{subequations}
In terms of the parameter $\rho$, we have
\begin{equation} \label{phit}
\phi_T =-\sqrt{\frac{4 (A^2+ K_0 e^{\rho}) (R^2- K_0 e^{\rho})}{K_0 e^{\rho} R^2 (A^2+2 K_0 e^{\rho})}},
\end{equation}
and (\ref{tol4mv}) corresponds to
\begin{subequations}\label{tol4nmv}
\begin{align}
\mathcal{K}_T(\rho) =&\frac{R^2 \left(A^2+2 K_0 e^{\rho}\right)}{4 \left(A^2+ K_0 e^{\rho}\right)
 \left(R^2- K_0 e^{\rho}\right)},\\
P_T(\rho) =& \frac{ K_0 e^{\rho} \left(A^2+3 K_0 e^{\rho}-R^2\right)}{4 \left(A^2+ K_0 e^{\rho}\right) \left( K_0 e^{\rho}-R^2\right)}, \\
\mathbb{M}_T(\rho) =& \frac{A^2 \left(2 A^2+R^2\right)}{4
 \left(A^2+R^2\right) \left(A^2+ K_0 e^{\rho }\right)}-\frac{A^2}{2 \left(A^2+2 K_0 e^{\rho}\right)}
 \nonumber\\
 &+\frac{R^2 \left(3 A^2+4R^2\right)}{4 \left(A^2+R^2\right) \left(R^2- K_0 e^{\rho}\right)}-\frac{3}{4}, \\
Y_T(\rho) =& \frac{K_0 e^{\rho}}{2 \left(A^2+ K_0 e^{\rho}\right)}.
\end{align}
\end{subequations}
%%%%%%%%%%%%%%%%%%%%%%%%%%%%%%%%%%%%%%%%%
\section{A direct resolution strategy}\label{RS}
%%%%%%%%%%%%%%%%%%%%%%%%%%%%%%%%%%%%%%%%%
In this section, we present a direct method to obtain two-fluid solutions from the single fluid ones given in Sec. \ref{KS}. We assume that the total isotropic and anisotropic pressures, density, and metric coefficients are those of known single fluid solutions given in Sec. \ref{KS}. Therefore, the spacetime of the objects described by these new solutions is that of the corresponding known single fluid solution. Using some additional conditions, we can then deduce the thermodynamical properties of a set of two fluids, thus obtaining a complete solution.  

In the following, we will present two-fluid solutions which satisfy the conditions in Sec. \ref{CON} for at least one set of parameters. Note that, for simplicity sake, we set $K_0$ and $r_0$ to unity from this point forward.
%%%%%%%%%%%%%%%%%%%%%%%%%%%%%%%%%%%%%%%%%%%%%%%%%%%%%%%%%%%%%%%%%%%%%%
\subsection{A two-fluid Bowers-Liang solution} \label{BL}
%%%%%%%%%%%%%%%%%%%%%%%%%%%%%%%%%%%%%%%%%%%%%%%%%%%%%%%%%%%%%%%%%%%%%%%
A first example is based on the idea behind the derivation of the Bowers-Liang solution \cite{bowers}. This is a widely used approach to find single fluid solutions \cite{bowers,herrera}, which involves making an ansatz on the behaviour of the anisotropy and solving for the corresponding pressure. 

We choose the total anisotropic pressure $\mathbb{P}_{tot}$ as the Bowers-Liang anisotropic pressure for a single fluid \eqref{ppbl}: 
\begin{equation}
    \mathbb{P}_{tot} = -\frac{e^{2 \rho} \mu_{tot}^2 }{2 z_1^2} \left[3^{\frac{h}{2}} (h-1)\mathfrak{c} z^{\frac{h}{2}-2}  \right] \label{pptbl},
\end{equation}
where $\mathfrak{c}$ is a generic constant, and we have used \eqref{z0} and \eqref{z1}. In this case, since we are extending the Bowers-Liang solution, $\mu_1$ and $\mu_2$ are constant. We can then write, from the definition of $\mathcal{K}$ and $\mathbb{M}_i$
\begin{subequations}\label{m1m2}
\begin{align}
\mathbb{M}_1 &= e^{\rho} \mu_1 \mathcal{K} \label{m1bl}, \\
\mathbb{M}_2 &= e^{\rho} \mu_2 \mathcal{K}.
\end{align}
\end{subequations}
It is easy to check that with the above choices \eqref{tovk} gives
\begin{equation}\label{K}
\mathcal{K} = \frac{3}{4 z}=\mathcal{K}_{BL},
\end{equation}
and substituting these results in \eqref{EqPtot} we have
\begin{equation}
P_{tot} = \frac{e^{\rho} \mu_{tot}}{4 z^2 z_1^2}  \left[z_2-3 z \left(3^{h+1} \mathfrak{c}^2+z^h\right)\right]=P_{BL}, \label{ptbl}
\end{equation}
as expected, where
\begin{equation}
z_2 = 2 \times 3^{h/2} \mathfrak{c} z^{h/2} \left((h+5) e^{\rho} \mu_{tot}-18\right).
\end{equation}
To solve for the individual fluids we write
\begin{subequations}\label{p1p2}
\begin{align}
P_{tot} = P_1 + P_2, \\
\mathbb{P}_{tot} = \mathbb{P}_1 + \mathbb{P}_2.
\end{align}
\end{subequations}
Let us now assume $\mathbb{P}_1 = 0$. Then $\mathbb{P}_2 = \mathbb{P}_{tot}$ and we can replace $P_2$ in the TOV equation for fluid 1 \eqref{p1} with $P_{tot} - P_2$. With $\mathbb{M}_1$ from \eqref{m1bl}, $\mathcal{K}$ from \eqref{K} and $\mathbb{P}_{tot}$ from \eqref{pptbl}, we can solve the resulting TOV equation for $P_1$. Subtraction from the total isotropic pressure $P$ gives $P_2$. 

\begin{figure} 
\centering
\includegraphics[scale=0.65]{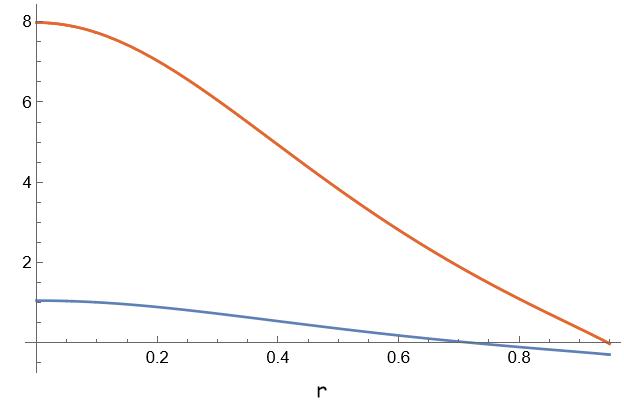}
\caption{Radial pressure for the Bowers-Liang solution in \eqref{bl} for fluid 1 (blue) and fluid 2 (green) and the tangential pressure for fluid 2 (red) vs $r$ with $h = 1$, $\mu_1 = 0.68$, $\mu_2 = 2.33$, $c_1 = -0.45$ and $\mathfrak{c} = -0.4$. The red and green curves overlap.}
\label{f1h1}
\end{figure}

\begin{figure} 
\centering
\includegraphics[scale=0.65]{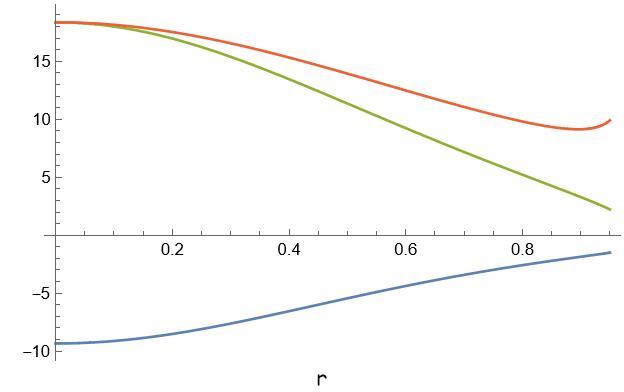}
\caption{Radial pressure for the Bowers-Liang solution in \eqref{bl} for fluid 1 (blue) and fluid 2 (green) and the tangential pressure for fluid 2 (red) vs $r$ with $h = \frac{1}{2}$, $\mu_1 = 0.68$, $\mu_2 = 2.33$, $c_1 = -0.45$ and $\mathfrak{c} = -0.4$.}
\label{f1h}
\end{figure}

\begin{figure} 
\centering
\includegraphics[scale=0.65]{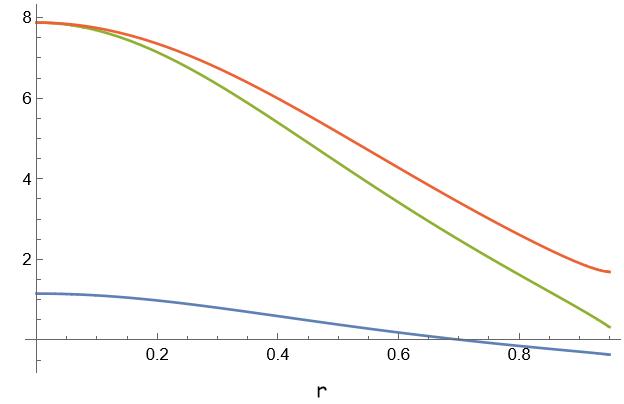}
\caption{Radial pressure for the Bowers-Liang solution in \eqref{bl} for fluid 1 (blue) and fluid 2 (green) and the tangential pressure for fluid 2 (red) vs $r$ with $h = 0.8$, $\mu_1 = 0.68$, $\mu_2 = 2.33$, $c_1 = -0.45$ and $\mathfrak{c} = -0.4$.}
\label{f1h8}
\end{figure}

\begin{figure} 
\centering
\includegraphics[scale=0.65]{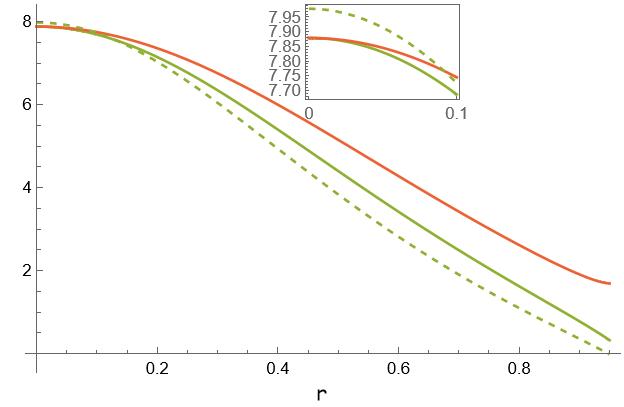}
\caption{Radial pressure for the Bowers-Liang solution in \eqref{bl} for fluid 2  with $h=0.8$ (solid green) and $h=1$ (dashed green) and the tangential pressure for fluid 2 (red) vs $r$ with $\mu_1 = 0.68$, $\mu_2 = 2.33$, $c_1 = -0.45$ and $\mathfrak{c} = -0.4$, zoomed in for $r = 0$ to $r = 0.1$.}
\label{f1hz}
\end{figure}

Using $\phi$ from \eqref{phibl} and \eqref{newvars}, as well as \eqref{prpt}, we can rewrite $P$ and $\mathbb{P}$ for each fluid in terms of $r$ and derive the radial and tangential pressures as 
\begin{subequations}\label{bl}
\begin{align}
p_{r1} = & p_{\perp 1} = -\mu_1+\frac{4}{3} c_1 \mathfrak{z}_1^{-\frac{1}{h}}, \\
p_{r2} = & \frac{2 \times 3^{\frac{h}{2}} \mu_{tot} \mathfrak{c}}{3^{\frac{h}{2}+1} \mathfrak{c}+\mathfrak{z}^{\frac{h}{2}}}-\frac{4}{3} c_1 \mathfrak{z}_1^{-\frac{1}{h}}-\mu_2,\\
p_{\perp 2} = & \frac{\mathfrak{z} \left(-3^{h+2} \mu_{tot} \mathfrak{c}^2-4 c_1 \mathfrak{z}_1^{2-\frac{1}{h}}+3 \mu_1 \mathfrak{z}_1^2\right)}{3 \mathfrak{z} \mathfrak{z}_1^2} \nonumber \\
& +\frac{3^{\frac{h}{2}+1} \mu_{tot} \mathfrak{c} \mathfrak{z}^{\frac{h}{2}} \left[(h+3) \mu_{tot} r^2-12\right]}{3 \mathfrak{z} \mathfrak{z}_1^2} \nonumber \\
&- \frac{3 \mu_{tot} \mathfrak{z}^{h+1}}{3 \mathfrak{z} \mathfrak{z}_1^2},
\end{align}
\end{subequations}
where
\begin{subequations}\label{zir}
\begin{align}
\mathfrak{z} = & 3-r^2 (\mu_1+\mu_2),\\
\mathfrak{z}_1 = & 3^{\frac{h}{2}+1} \mathfrak{c}+\mathfrak{z}^{\frac{h}{2}}.
\end{align}
\end{subequations}
The above solution is plotted in Figs. \ref{f1h1}, \ref{f1h} and \ref{f1h8} for three different values of $h$. Figure \ref{f1h1} shows the case $h=1$ for which both fluids are isotropic. For the same set of parameters we can plot solution \eqref{bl} for $h= \frac{1}{2}$. It is evident from Fig. \ref{f1h} that in this case, the solution is not physical, as the radial pressure of fluid 1 is negative. This result shows that not all values of $h$ are compatible with a given set of parameters. If we take instead $h=0.8$ (see Fig. \ref{f1h8}), the solution is again physical and we can estimate the effect that the presence of anisotropy has on the thermodynamical quantities. In fact,  Fig. \ref{f1hz} shows that the anisotropy lowers the radial pressure at the core and raises the radial pressure at the surface, suggesting that anisotropic compact stellar objects can have different structures with respect to their isotropic counterparts. This is consistent with the results of \cite{viaggiu} already mentioned in the introduction. 

Finally, as in \eqref{bl} the radial  pressure of fluid 1 goes to zero before that of fluid 2, this solution can only represent a shell and will have to be joined to another solution to  describe a complete compact stellar object.

%%%%%%%%%%%%%%%%%%%%%%%%%%%%%%%%%%%%%%%%%%%%%%%%%%%%%%%%%%%%%%%%%%%%%%%%
\subsection{Two Florides-class solutions}\label{FL1}
%%%%%%%%%%%%%%%%%%%%%%%%%%%%%%%%%%%%%%%%%%%%%%%%%%%%%%%%%%%%%%%%%%%%%%%%
We utilize the single fluid Florides-class constant density solution in Sec. \ref{Flor}. For these solutions, the relation \eqref{fpp} always holds. For notational convenience, we will write this relation as
\begin{equation}\label{fpp_tot}
\mathbb{P}_{tot} = \frac{1}{6}  \mathbb{M}_{tot} (1-4 \mathcal{K}_F).
\end{equation}
Following \cite{sante2}, using the assumption above and defining the quantity  
\begin{equation}\label{fp}
\mathcal{P}_{tot} = P_{tot} + \mathbb{P}_{tot},
\end{equation}
related with the radial pressure, one can write the total TOV equation as
\begin{equation}\label{Fl_Eq}
\mathcal{P}_{tot,\rho} + \mathcal{P}_{tot}^2+\mathcal{P}_{tot}\left(3 \mathcal{K}-\mathbb{M}_1-\mathbb{M}_2-\frac{7}{4}\right)=0,
\end{equation}
where $\mathcal{K}=\mathcal{K}_F$ is given by \eqref{K_F} [or can derived by \eqref{tovk}, remembering that in this case  $\mathbb{M}_1 $ and $\mathbb{M}_2 $ are given by \eqref{m1m2}]. The solution of Eq. \eqref{Fl_Eq} reads  
\begin{equation}
\mathcal{P}_{tot} = \frac{1}{e^{\mathfrak{c}-\rho}+1}. \label{fcalptot}
\end{equation}
Now Eq. \eqref{fp} can be used to determine $P_{tot}$ which reads
\begin{equation}
P_{tot} = \frac{e^{2 \rho} \mu_{tot}^2}{8 z^2}+\frac{1}{e^{\mathfrak{c}-\rho}+1},
\end{equation}
and \eqref{fpp_tot} gives
\begin{equation}
\mathbb{P}_{tot} = -\frac{e^{2 \rho} \mu_{tot}^2}{8 z^2}.
\end{equation}

In order to find the pressures of the two fluids of the new solution we can use the relations
\begin{eqnarray}
\mathcal{P}_i &=& P_i + \mathbb{P}_i, \qquad i=1,2 \label{pcurlyp}\\
P_2 &=& P_{tot} - P_1, \label{ptotp}\\
\mathcal{P}_2 &=& \mathcal{P}_{tot} - \mathcal{P}_1,
\end{eqnarray}
in the TOV equation for fluid 1 \eqref{p1} so that it will contain only $\mathcal{P}_{1,\rho}$, $\mathbb{P}_1$, $\mathcal{P}_{tot}$, $\mathbb{M}_1$ and $\mathbb{M}_2$:
\begin{eqnarray}\label{fcalp} 
\mathcal{P}_{1,\rho} = &&\mathcal{P}_1 \left[\frac{7}{4} -3 \mathcal{K}+2 - \mathcal{P}_{tot} +2 \mathbb{M}_{tot}\right] \nonumber \\
&&+\frac{1}{4} \mathbb{M}_1 (-4 \mathcal{K}-4 \mathcal{P}_{tot}+1)-\frac{3}{2}\mathbb{P}_1,
\end{eqnarray}
where $\mathbb{M}_{tot} = \mathbb{M}_1 + \mathbb{M}_2$ and $\mathcal{K}=\mathcal{K}_F$.

We now consider two cases: The first, in which one of the fluids is isotropic $(\mathbb{P}_1 = 0)$ and a second, in which both fluids are anisotropic. For convenience we will suppose, without loss of generality, that fluid 1 changes its character.

\subsubsection*{The case $\mathbb{P}_1 = 0$}
We take $\mathbb{P}_1 = 0$, so that $\mathcal{P}_{1}={P}_{1}$ and $\mathbb{P}_2 = \mathbb{P}_{tot}$, and the TOV equation for the first fluid \eqref{fcalp} becomes
\begin{eqnarray}\label{fcalp1} 
\mathcal{P}_{1,\rho} = &&\mathcal{P}_1 \left[\frac{7}{4} -3 \mathcal{K}+2 - \mathcal{P}_{tot} +2 \mathbb{M}_{tot}\right] \nonumber \\
&&+\frac{1}{4} \mathbb{M}_1 (-4 \mathcal{K}-4 \mathcal{P}_{tot}+1).
\end{eqnarray}

\begin{figure} 
\centering
\includegraphics[scale=0.65]{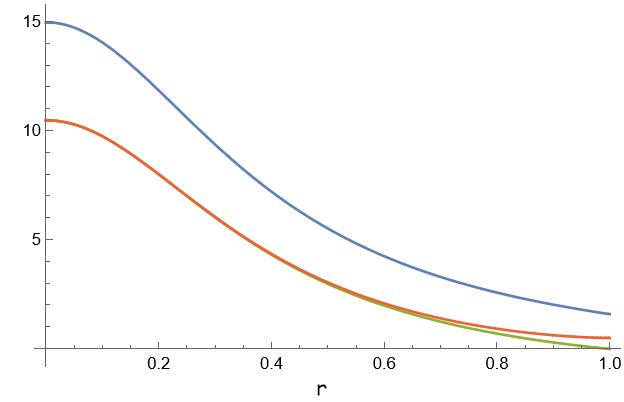}
\caption{Radial pressure for the constant density Florides-class solution in \eqref{f} for fluid 1 (blue) and fluid 2 (green), and tangential pressure for fluid 2 (red) vs $r$ with $\mu_1 = 0.1$, $\mu_2 = 1.55$, $c_1 = 1.35$ and $\mathfrak{c} = -1.85$.}
\label{f2}
\end{figure}

We can solve this equation for $\mathcal{P}_1$ with \eqref{m1m2}, \eqref{K} and \eqref{fcalptot}, and obtain $\mathcal{P}_2$ by subtracting from the total $\mathcal{P}_{tot}$. Since $P_1 = \mathcal{P}_1$, we obtain $P_2$ by subtracting $\mathcal{P}_1$ from $P_{tot}$. As $\mathbb{P}_2 = \mathbb{P}_{tot}$, all the pressures are known in terms of $\rho$ and we can utilize $\phi$ from \eqref{phif} with \eqref{newvars} to give $p_1, p_2, \Pi_1,$ and $\Pi_2$ in terms of $r$. Finally, using \eqref{prpt},  we can write the radial and tangential pressures for each fluid in terms of $r$ as
\begin{subequations}\label{f}
\begin{align}
p_{r1} = & p_{\perp 1} = -\mu_1+\frac{4 c_1 \mathfrak{z}^{\frac{1}{4}}}{3 \left(e^{\mathfrak{c}}+r^2\right)},\\
p_{r2} = & \frac{3 \mu_1 e^{\mathfrak{c}}-r^2 (\mu_1+4 \mu_2)-4 c_1 \mathfrak{z}^{\frac{1}{4}}+12}{3 \left(e^{\mathfrak{c}}+r^2\right)},\\
p_{\perp 2} = & \frac{9 \mu_1-3 \mu_2+3 \mu_{tot}+\mu_{tot} r^2 (\mu_2-3 \mu_1)}{12-4 \mu_{tot} r^2} \nonumber \\
&-\frac{4 \left(\mu_{tot} r^2+c_1 \mathfrak{z}^{\frac{1}{4}}-3\right)}{3 \left(e^{\mathfrak{c}}+r^2\right)},
\end{align}
\end{subequations}
where we have used the $\mathfrak{z}_i$ from \eqref{zir}. This solution is plotted in Fig. \ref{f2}, and represents a shell as the radial pressure of fluid 2 approaches zero before that of fluid 1.
%%%%%%%%%%%%%%%%%%%%%%%%%%%%%%%%%%%%%%%%%%%%%%%%%%%%%%%%%%%%%%%%%%%
\subsubsection*{The case $\mathbb{P}_1 \neq 0$}
%%%%%%%%%%%%%%%%%%%%%%%%%%%%%%%%%%%%%%%%%%%%%%%%%%%%%%%%%%%%%%%%%%%
When fluid 1 is anisotropic, we need an ansatz for  $\mathbb{P}_1$  in order to solve \eqref{fcalp}. One convenient choice is
\begin{equation} \label{p1constr}
\mathbb{P}_1 = \frac{1}{6}  \mathbb{M}_1 (1-4 \mathcal{K}_F).
\end{equation}
In this way, equation \eqref{fcalp} takes the form
 \begin{equation}\label{fcalp2} 
\mathcal{P}_{1,\rho} = \mathcal{P}_1 \left(-3 \mathcal{K}_F + 2\mathbb{M}_{tot}-\mathcal{P}_{tot}+\frac{7}{4}\right)-\mathbb{M}_1 \mathcal{P}_{tot}.
\end{equation}

As before, once we know $\mathcal{P}_1$ we can obtain $\mathcal{P}_2$ by subtraction from $\mathcal{P}_{tot}$ and ${P}_1$ by subtraction of $\mathbb{P}_{1}$ in \eqref{p1constr} from $\mathcal{P}_{1}$. Finally $\mathcal{P}_2$, $\mathbb{P}_2$ and $P_2$ are derived from the equations for the total pressures. 

\begin{figure} 
\centering
\includegraphics[scale=0.65]{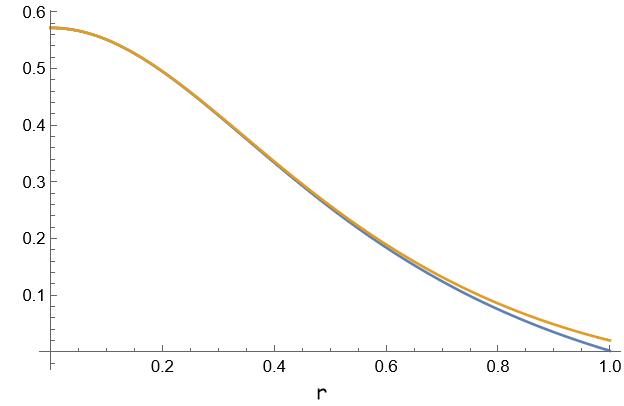}
\caption{Radial pressure (blue) and tangential pressure (orange) for the constant density Florides-class solution in \eqref{f2a} for fluid 1 vs $r$ with $\mu_1 = 0.2$, $\mu_2 = 0.6$, $c_1 = -1$, $c_2 = -0.45$ and $\mathfrak{c} = -0.05$.}
\label{f3}
\end{figure}

\begin{figure} 
\centering
\includegraphics[scale=0.65]{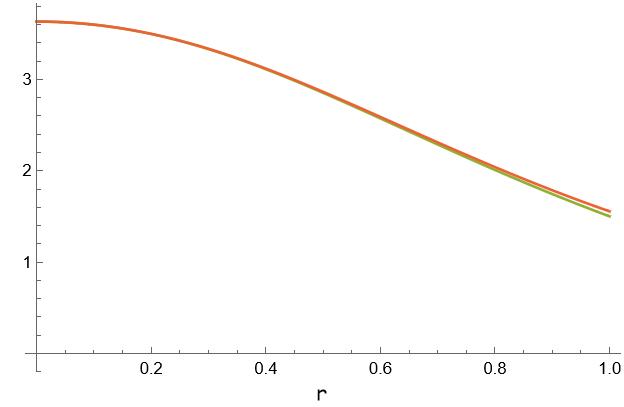}
\caption{Radial pressure (green) and tangential pressure (red) for the constant density Florides-class solution in \eqref{f2a} for fluid 2 vs $r$ with $\mu_1 = 0.2$, $\mu_2 = 0.6$, $c_1 = -1$, $c_2 = -0.45$ and $\mathfrak{c} = -0.05$. We have truncated this graph to the value of $r$ where the radial pressure of fluid 1 approaches zero.}
\label{f4}
\end{figure}

Once again, we utilize $\phi$ from \eqref{phif}, \eqref{newvars} to give $p_1, p_2, \Pi_1,$ and $\Pi_2$ in terms of $r$. With equation \eqref{prpt}, we can write the radial and tangential pressures for each fluid in terms of $r$ as
\begin{subequations}\label{f2a}
\begin{align}
p_{r1} = &\frac{4 \mu_1 \mathfrak{z}+4 c_2 \mu_{tot} \mathfrak{z}^{\frac{1}{4}}}{3 \mu_{tot} \left(r^2+e^{c_1}\right)},\\
p_{\perp 1} = &\frac{\mu_1 \mu_{tot} r^2}{4 \mathfrak{z}}+\frac{4 \left(\mu_1 \mathfrak{z}+c_2 \mu_{tot} \mathfrak{z}^{\frac{1}{4}}\right)}{3 \mu_{tot} \left(r^2+e^{c_1}\right)},\\
p_{r2} = & \frac{4}{3} \left(\frac{\mathfrak{z}}{e^{\mathfrak{c}}+r^2}-\frac{\mu_1 \mathfrak{z}+c_2 \mu_{tot} \mathfrak{z}^{\frac{1}{4}}}{\mu_{tot} r^2+e^{c_1} \mu_{tot}}\right),\\
p_{\perp 2} = & \frac{4 \mathfrak{z}}{3 \left(e^{\mathfrak{c}}+r^2\right)}-\frac{4 \mathfrak{z} \left(\frac{\mu_1}{\mu_{tot}}+\frac{c_2}{\mathfrak{z}^{3/4}}\right)}{3 \left(r^2+e^{c_1}\right)}+\frac{\mu_2 \mu_{tot} r^2}{4 \mathfrak{z}},
\end{align}
\end{subequations}
where $c_1$, $c_2$ and $\mathfrak{c}$ are constants. This solution is plotted in Figs. \ref{f3} and \ref{f4}. We can see that for this solution, the radial pressure of fluid 1 approaches zero before fluid 2, so this solution represents a shell.
%%%%%%%%%%%%%%%%%%%%%%%%%%%%%%%%%%%%%%%%%%%%%%%%%%%%%%%%%%%%%%%%%%%%%%%%%%%%
%%%%%%%%%%%%%%%%%%%%%%%%%%%%%%%%%%%%%%%%%%%%%%%%%%%%%%%%%%%%%%%%%%%%%%%%%%%
%%%%%%%%%%%%%%%%%%%%%%%%%%%%%%%%%%%%%%%%%%%%%%%%%%%%%%%%%%%%%%%%%%%%%%%%
\subsection{Two Florides-class solutions with fractional density}\label{FL2}
%%%%%%%%%%%%%%%%%%%%%%%%%%%%%%%%%%%%%%%%%%%%%%%%%%%%%%%%%%%%%%%%%%%%%%%%
 We can consider a variation of the previous strategy assuming a form for the pressure $\mathcal{P}_{tot}$ and the energy density. We choose the total pressure variable $\mathcal{P}_{tot}$ as 
\begin{equation} \label{pr}
\mathcal{P}_{tot} = -\frac{3 e^{\rho} \left(b e^{\rho} -a\right)}{z_3},
\end{equation}
where $a$ and $b$ are constants and 
\begin{equation}
    z_3 =12 a^2-3 a (4 b-1) e^{\rho}+2 b (2 b-1) e^{2 \rho}.
\end{equation} 
We can then derive $\mathbb{M}_{tot}$ from Eq. \eqref{Fl_Eq} and use this quantity to find $\mathbb{P}_{tot}$ via \eqref{fpp_tot} and solve for $\mathcal{K}$ via \eqref{tovk}. At this point, using \eqref{fp} we can easily obtain $P_{tot}$. Utilizing this procedure, we have
\begin{subequations} \label{fptotsfd}
\begin{align}
P_{tot} = &\frac{e^{\rho} z_6}{z_4}, \\
\mathbb{P}_{tot} = &\frac{ e^{2 \rho}}{z_4} \left(4 b+1\right)^2 \left(2 b e^{\rho} -3 a\right)z_5,
\end{align}
\end{subequations}
where
\begin{subequations}
\begin{align}
z_4 &=24 \left(a-b e^{\rho} \right) z_3^2,\\
z_5 &= 9 a^2-7 a b e^{\rho} +2 b^2 e^{2 \rho},\\
z_6 &= -3 a^2 b \left[8 b (26 b-107) +205 \right] e^{2 \rho}\nonumber \\
&+27 a^3 \left[8 b (2 b-11)+9 \right] e^{\rho} +864 a^4 \nonumber \\
&+4 a b^2 \left[80 (b-4) b+131 \right] e^{3 \rho}-4 b^3 \left[16 (b-4) b \right. \nonumber \\
& \left. +37 \right] e^{4 \rho}.
\end{align}
\end{subequations}

We now need to give an ansatz for the energy density of the single fluids. In fact, our choice of $\mathcal{P}_{tot} $ can only determine, via the equation of state, the total density.

As an example, we will set $\mathbb{M}_1 = \frac{1}{4} \mathbb{M}_{tot}$, such that $\mathbb{M}_2 = \frac{3}{4} \mathbb{M}_{tot}$, where
\begin{equation}\label{fmt}
\mathbb{M}_{tot} = -\frac{ e^{\rho} z_4 z_5}{6 z_3} (4 b+1),
\end{equation}
and
\begin{equation} \label{fKfd}
\mathcal{K} = \frac{3 \left(a-b e^{\rho} \right)^2}{z_3}.
\end{equation}

As before, we consider  two cases. The first in which one of the fluids is isotropic $(\mathbb{P}_1 = 0)$ and a second in which both fluids are anisotropic. 
%%%%%%%%%%%%%%%%%%%%%%%%%%%%%%%%%%%%%%%%%%%%%%
\subsubsection*{The case $\mathbb{P}_1 = 0$}
\begin{figure} 
\centering
\includegraphics[scale=0.65]{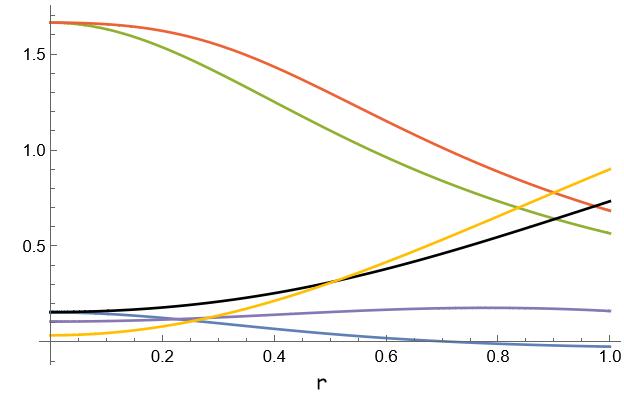}
\caption{Radial pressure for the Florides-class solution in \eqref{f3a} for fluid 1 (blue) and fluid 2 (green), tangential pressure for fluid 2 (red), $dp_r/d\mu$ for fluid 1 (purple) and fluid 2 (black), and $dp_{\perp}/d\mu$ for fluid 2 (yellow) vs $r$ with $a = 0.55$, $b = -1.3$, and $c_1 = -1.3$.}
\label{f5a}
\end{figure}

\begin{figure} 
\centering
\includegraphics[scale=0.65]{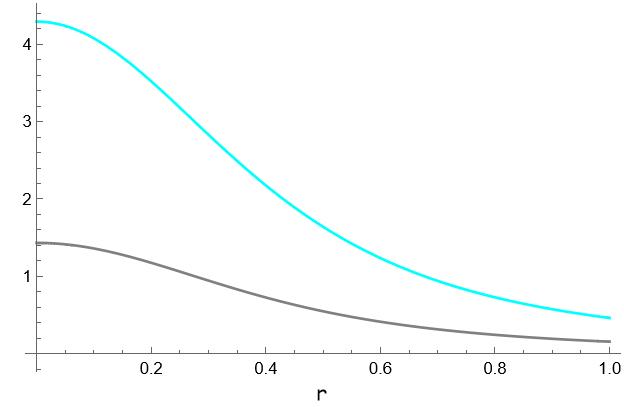}
\caption{Energy density profile for the Florides-class solution in \eqref{f3d} for fluid 1 (gray) and fluid 2 (cyan) vs $r$ with $a = 0.55$, $b = -1.3$, and $c_1 = -1.3$.}
\label{f5d}
\end{figure}

%%%%%%%%%%%%%%%%%%%%%%%%%%%%%%%%%%%%%%%%%%%%%%
In the case $\mathbb{P}_1 = 0$, we have $\mathbb{P}_2 = \mathbb{P}_{tot}$ and $\mathcal{P}_1 = P_1$. With \eqref{fptotsfd}, \eqref{fmt} and \eqref{fKfd}, we can solve the resulting TOV equation for fluid 1 \eqref{fcalp} with \eqref{K}, \eqref{fcalptot}, and our ansatz on $\mathbb{M}_1$. We can obtain $\mathcal{P}_2$ by subtracting from the total $\mathcal{P}_{tot}$. Since $P_1 = \mathcal{P}_1$, we obtain $P_2$ by subtracting from $P_{tot}$. As $\mathbb{P}_2 = \mathbb{P}_{tot}$, we proceed to utilize $\phi$ from \eqref{phif} with \eqref{newvars} to give $p_1, p_2, \Pi_1,$ and $\Pi_2$ in terms of $r$. Finally, using \eqref{prpt}, we can write the radial and tangential pressures for each fluid in terms of $r$ as

\begin{subequations}\label{f3a}
\begin{align}
p_{r1} =  p_{\perp 1}=& \frac{1}{4 a-4 b r^2} +\frac{1}{3} e^{\alpha } c_1 \left[12 a^2  +3 a (1-4 b) r^2\right. \nonumber \\
&\left. +2 b (2 b-1) r^4\right]^{\frac{5-4 b}{16 b-8}},\\
p_{r2} = & \frac{3}{4 \left(a-b r^2\right)}-\frac{1}{3} e^{\alpha } c_1 \left[12 a^2+3 a (1-4 b) r^2\right. \nonumber \\
&\left. +2 b (2 b-1) r^4\right]^{\frac{5-4 b}{16 b-8}},\\
p_{\perp 2} = & \mathfrak{z}_2 -\frac{1}{3} e^{\alpha } c_1 \left[12 a^2+3 a (1-4 b) r^2\right. \nonumber \\
&\left. +2 b (2 b-1) r^4\right]^{\frac{5-4 b}{16 b-8}},
\end{align}
\end{subequations}
and the energy densities of both fluids as 
\begin{subequations}\label{f3d}
\begin{align}
\mu_{1} =& -\frac{(4 b+1) \left(9 a^2-7 a b r^2+2 b^2 r^4\right)}{48 \left(a-b r^2\right)^3},\\
\mu_{2} = & -\frac{(4 b+1) \left(9 a^2-7 a b r^2+2 b^2 r^4\right)}{16 \left(a-b r^2\right)^3},
\end{align}
\end{subequations}

where 
\begin{subequations}
\begin{align}
\alpha =& \frac{\sqrt{48 b^2-24 b-9} \tan ^{-1}\left[\frac{a (3-12 b)+4 b (2 b-1) r^2}{a \sqrt{48 b^2-24 b-9}}\right]}{8 b (4 b-5)+12},\\
\mathfrak{z}_2 = & \left\{-3 a^2 b \left[8 b (26 b-47)+109\right] r^4+27 a^3 \left[8 b (2 b-5)\right. \right. \nonumber \\
&\left. \left. +5\right] r^2+432 a^4+16 a b^2 \left[5 b (4 b-7)+17\right] r^6 \right. \nonumber \\
&\left. -4 b^3 \left[4 b (4 b-7)+19\right] r^8\right\} \times \left\{48 \left(a-b r^2\right)^3 \right. \nonumber \\
&\left. \left[12 a^2+3 a (1-4 b) r^2+2 b (2 b-1) r^4\right]\right\}^{-1}.
\end{align}
\end{subequations}
This solution, which only describes a shell, is plotted in Figs. \ref{f5a} and \ref{f5d}.
%%%%%%%%%%%%%%%%%%%%%%%%%%%%%%%%%%%%%%%%%%%%%%%%%%%%%%%%%%%%%%%%%%%%%%%%%%%%%%
%%%%%%%%%%%%%%%%%%%%%%%%%%%%%%%%%%%%%%%%%%%%%%%%%%%%%%%%%%%%%%%%%%%%%%%%%%%%%%%
%%%%%%%%%%%%%%%%%%%%%%%%%%%%%%%%%%%%%%%%%%%%%%%%%%%%%%%%%%%%%%%%%%%%%%%%%%%%%%%%%%%%
\subsubsection*{The case $\mathbb{P}_1 \neq 0$}
%%%%%%%%%%%%%%%%%%%%%%%%%%%%%%%%%%%%%%%%%%%%%%%%%%%%%%%%%%%%%%%%%%%%%%%%%%%%%%%%%%%%%
We proceed as in the $\mathbb{P}_1 = 0$ case, and assume again \eqref{p1constr} in \eqref{fcalp} to obtain the resulting TOV equation \eqref{fcalp2}, which we can solve for $\mathcal{P}_1$ with \eqref{m1m2}, \eqref{K}, and \eqref{fcalptot}. We have $\mathbb{P}_1$ from the relation \eqref{p1constr}, and $P_1$ is given in \eqref{pcurlyp}. We can obtain $\mathcal{P}_2$, $\mathbb{P}_2$ and $P_2$ by subtracting from the total pressures. Next, we utilize $\phi$ from \eqref{phif} with \eqref{newvars} to give $p_1, p_2, \Pi_1,$ and $\Pi_2$ in terms of $r$. With equation \eqref{prpt}, we can write the radial and tangential pressures for each fluid in terms of $r$. The quantities $p_{r1}, p_{\perp 1}, p_{r2},$ and $p_{\perp 2}$ are too long to be presented here, and are  given by \eqref{f4a} in Appendix \ref{AARS}. 

\begin{figure} 
\centering
\includegraphics[scale=0.65]{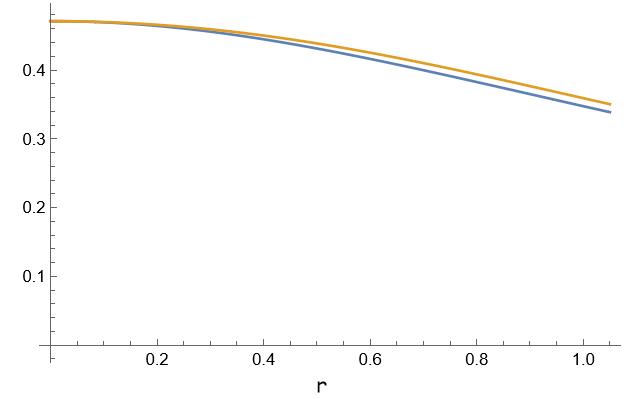}
\caption{Radial pressure (blue) and tangential pressure (orange) for the Florides-class solution in \eqref{f4a} for fluid 1 vs $r$ with $a = 1.75$, $b = -1.1$, and $c_1 = 3.35$. As the radial pressure for fluid 2 approaches zero before fluid 1, we have truncated the graph for fluid 1.}
\label{f6}
\end{figure}

\begin{figure} 
\centering
\includegraphics[scale=0.65]{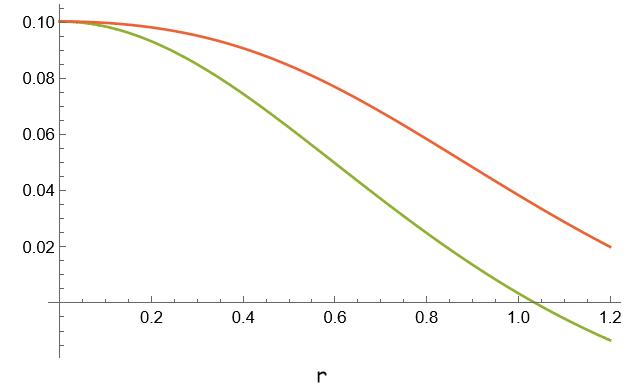}
\caption{Radial pressure (green) and tangential pressure (red) for the Florides-class solution in \eqref{f4a} for fluid 2 vs $r$ with $a = 1.75$, $b = -1.1$, and $c_1 = 3.35$.}
\label{f7}
\end{figure}

\begin{figure} 
\centering
\includegraphics[scale=0.65]{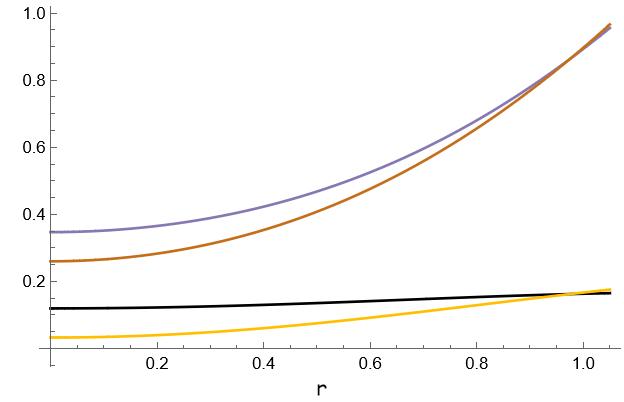}
\caption{$dp_r/d\mu$ for fluid 1 (purple) and fluid 2 (brown), and $dp_{\perp}/d\mu$ for fluid 1 (black) and fluid 2 (yellow) for the Florides-class solution in \eqref{f4a} for fluid 1 vs $r$ with $a = 1.75$, $b = -1.1$, and $c_1 = 3.35$.}
\label{f67a}
\end{figure}

\begin{figure} 
\centering
\includegraphics[scale=0.65]{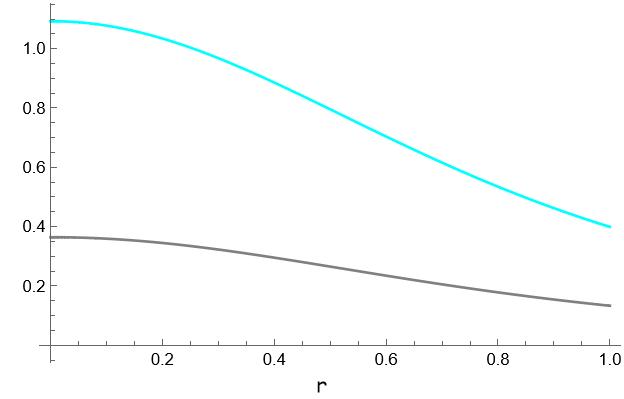}
\caption{Energy density profile for the Florides-class solution in \eqref{f7ed} for fluid 1 (gray) and fluid 2 (cyan) vs $r$ with $a = 1.75$, $b = -1.1$, and $c_1 = 3.35$.}
\label{f7d}
\end{figure}

The energy densities of both fluids are 
\begin{subequations}\label{f7ed}
\begin{align}
\mu_{1} =& -\frac{(4 b+1) \left(9 a^2-7 a b r^2+2 b^2 r^4\right)}{48 \left(a-b r^2\right)^3},\\
\mu_{2} = & -\frac{(4 b+1) \left(9 a^2-7 a b r^2+2 b^2 r^4\right)}{16 \left(a-b r^2\right)^3},
\end{align}
\end{subequations}

and are shown in Fig. \ref{f7d}. The solution represented by \eqref{f4a} is plotted in Fig. \ref{f6}, \ref{f7} and \ref{f67a}. We observe that the radial pressure of fluid 2 approaches zero before that of fluid 1. Thus, this solution represents a shell.
%%%%%%%%%%%%%%%%%%%%%%%%%%%%%%%%%%%%%%%%%%%%%%%%%%%%%%%%%%%%%%%%%%%%%%%%%%%%%%%%%%%%%%%
%%%%%%%%%%%%%%%%%%%%%%%%%%%%%%%%%%%%%%%%%%%%%%%%%%%%%%%%%%%%%%%%%%%%%%%%%%%%%%%%%%%%%%%
%%%%%%%%%%%%%%%%%%%%%%%%%%%%%%%%%%%%%%%%%%%%%%%%%%%%%%%%%%%%%%%%%%%%%%%%%%%%%%%%%%%%%%%
%%%%%%%%%%%%%%%%%%%%%%%%%%%%%%%%%%%%%%%%%%%%%%%%%%%%%%%%%%%%%%%%%%%%%%%%%%%%%%
%%%%%%%%%%%%%%%%%%%%%%%%%%%%%%%%%%%%%%%%%%%%%%%%%%%%%%%%%%%%%%%%%%%%%%%%%%%%%%%
\subsection{Two Florides-class solutions with barotropic equation of state}\label{FL3}
%%%%%%%%%%%%%%%%%%%%%%%%%%%%%%%%%%%%%%%%%%%%%%%%%%%%%%%%%%%%%%%%%%%%%%%%%%%%%%%
We follow the same procedure as in Sec. \ref{FL2}, with a different ansatz. Instead of setting $\mathbb{M}_1$ and $\mathbb{M}_2$ as fractions of $\mathbb{M}_{tot}$, we choose the equation of state. In particular we assume
\begin{equation}\label{eos}
\mathbb{M}_1 = \gamma \mathcal{P}_1.
\end{equation}
We now consider the usual two cases: one in which fluid 1 is isotropic  $(\mathbb{P}_1 = 0)$ and one in which both fluids are anisotropic. 

\subsubsection*{The case $\mathbb{P}_1 = 0$}
If we set the anisotropic pressure of one of the fluids to zero, $\mathbb{P}_{2}=\mathbb{P}_{tot}$ and the equation of state \eqref{eos} takes the form 
\begin{equation}
\mathbb{M}_1 = \gamma {P}_1.
\end{equation}
We can now solve the resulting TOV equation for fluid 1 \eqref{fcalp} with \eqref{fptotsfd}, \eqref{fmt}, \eqref{fKfd} and \eqref{eos}. Subtraction from the total pressures determines the pressures for fluid 2. 

As in the previous cases, we utilize $\phi$ from \eqref{phif} with \eqref{newvars} to give $p_1, p_2, \Pi_1,$ and $\Pi_2$ in terms of $r$. Finally, using \eqref{prpt}, we can write the radial and tangential pressures for each fluid in terms of $r$. Unfortunately, the full expressions for the exact solution represented by $p_{r1}, p_{\perp 1}, p_{r2},$ and $p_{\perp 2}$ are too long to be reported here. We refer the interested reader to \eqref{feos1a} of Appendix \ref{AARS}. 

The energy densities of both fluids are 
\begin{subequations}\label{fe1ad}
\begin{align}
\mu_{1} =& \frac{1}{3} \gamma  c_1 \left(12 a^2+3 a (1-4 b) r^2+2 b (2 b-1) r^4\right)^{\mathfrak{z}_3} \exp \beta,\\
\mu_{2} = & \frac{1}{12} \left(\frac{(4 b+1) \left(9 a^2-7 a b r^2+2 b^2 r^4\right)}{\left(b r^2-a\right)^3} \right.\nonumber\\
& \left. -4 \gamma  c_1 \left(12 a^2+3 a (1-4 b) r^2+2 b (2 b-1) r^4\right)^{\mathfrak{z}_3} \exp \beta \right),
\end{align}
\end{subequations}
where 
\begin{equation}\label{fz3}
\mathfrak{z}_3 = \frac{-4 b+6 \gamma +5}{16 b-8},
\end{equation}
and we have used \eqref{beta}. 

\begin{figure} 
\centering
\includegraphics[scale=0.65]{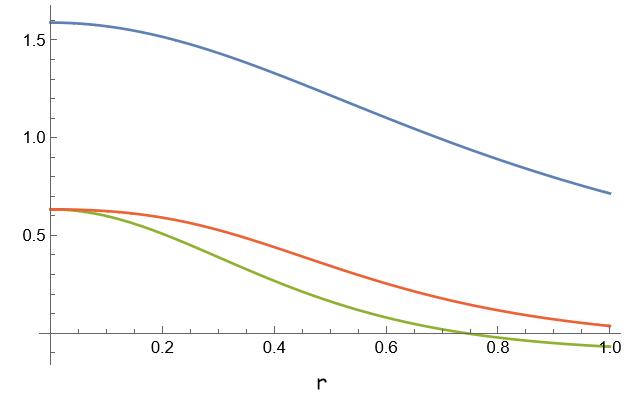}
\caption{Radial pressure for fluid 1 (blue) and fluid 2 (green), and tangential pressure for fluid 2 (red) for the Florides-class solution in \eqref{feos1a} vs $r$ with $a = 0.45$, $b = -1.1$, $c_1 = 6$ and $\gamma = 0.3$.}
\label{feos1}
\end{figure}

\begin{figure} 
\centering
\includegraphics[scale=0.65]{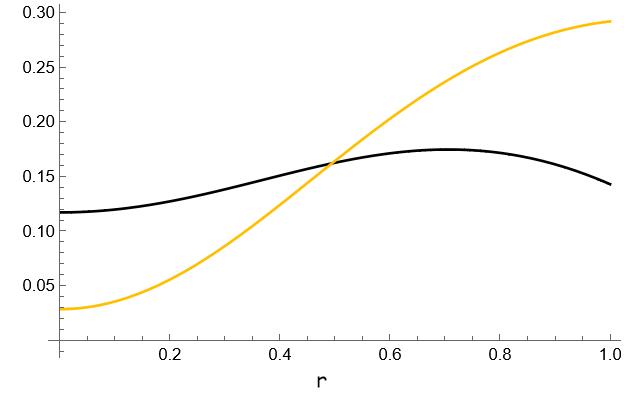}
\caption{$dp_r/d\mu$ (black) and $dp_{\perp}/d\mu$ (yellow) for fluid 2 for the Florides-class solution in \eqref{feos1a} vs $r$ with $a = 0.45$, $b = -1.1$, $c_1 = 6$ and $\gamma = 0.3$.}
\label{feos1af}
\end{figure}

\begin{figure} 
\centering
\includegraphics[scale=0.65]{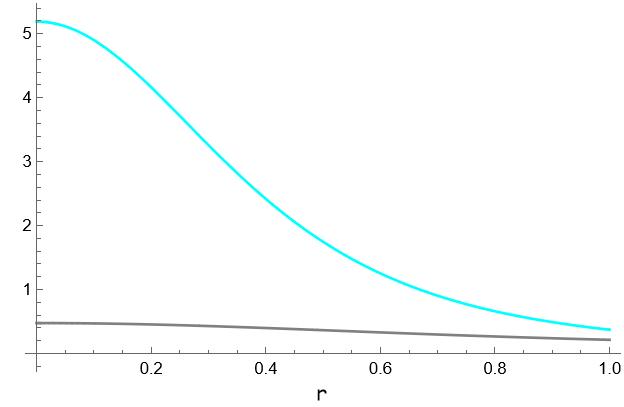}
\caption{Energy density profile for the Florides-class solution in \eqref{fe1ad} for fluid 1 (gray) and fluid 2 (cyan) vs $r$ with $a = 0.45$, $b = -1.1$, $c_1 = 6$ and $\gamma = 0.3$.}
\label{fe1d}
\end{figure}

The energy densities are shown in Fig. \ref{fe1d}. The pressures for both fluids are plotted in Figs. \ref{feos1} and \ref{feos1af}. The radial pressures of both fluids do not approach zero at the same value of $r$ as the radial pressure of fluid 2 approaches zero before that of fluid 1. Thus, this solution represents a shell. 
%%%%%%%%%%%%%%%%%%%%%%%%%%%%%%%%%%%%%%%%%%%%%%%%%%%%%%%%%%%%%%%%%%%%%%%%%%%%%%%
%%%%%%%%%%%%%%%%%%%%%%%%%%%%%%%%%%%%%%%%%%%%%%%%%%%%%%%%%%%%%%%%%%%%%%%%%%%%%%%
%%%%%%%%%%%%%%%%%%%%%%%%%%%%%%%%%%%%%%%%%%%%%%%%%%%%%%%%%%%%%%%%%%%%%%%%%%%%%%
%%%%%%%%%%%%%%%%%%%%%%%%%%%%%%%%%%%%%%%%%%%%%%%%%%%%%%%%%%%%%%%%%%%%%%%%%%%%%%%
\subsubsection*{The case $\mathbb{P}_1 \neq 0$}
%%%%%%%%%%%%%%%%%%%%%%%%%%%%%%%%%%%%%%%%%%%%%%%%%%%%%%%%%%%%%%%%%%%%%%%%%%%%%%%
In the general case with both fluids being anisotropic, we follow the same procedure as in the $\mathbb{P}_1 = 0$ case and use the equation of state \eqref{eos}. The relation \eqref{p1constr} then gives $\mathbb{P}_1$. Using \eqref{pcurlyp}, we can calculate $P_1$ from $\mathcal{P}_1$. The corresponding quantities for fluid 2 are calculated by subtracting from the total pressures. As usual, we utilize $\phi$ from \eqref{phif} with \eqref{newvars} to give $p_1, p_2, \Pi_1,$ and $\Pi_2$ in terms of $r$. Using \eqref{prpt}, we can write the radial and tangential pressures for each fluid in terms of $r$. The solution is too lengthy to include here, so we have listed it as \eqref{feos2a} in Appendix \ref{AARS}. 

\begin{figure} 
\centering
\includegraphics[scale=0.65]{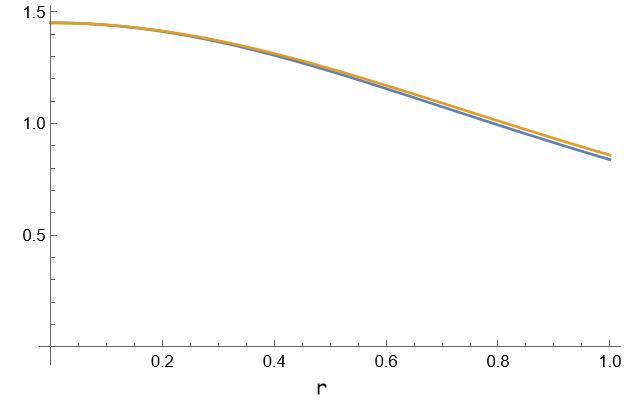}
\caption{Radial pressure (blue) and tangential pressure (orange) for the Florides-class solution in \eqref{feos2a} for fluid 1 vs $r$ with $a = 0.6$, $b = -0.6$, $\mathfrak{c}_1 = 7.8$ and $\gamma = 0.3$. This graph is truncated to the value of $r$ that the radial pressure of fluid 2 approaches zero.}
\label{feos2f1}
\end{figure}

\begin{figure} 
\centering
\includegraphics[scale=0.65]{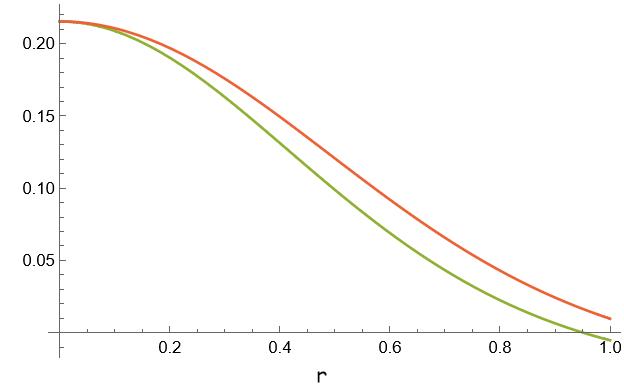}
\caption{Radial pressure (green) and tangential pressure (red) for the Florides-class solution in \eqref{feos2a} for fluid 2 vs $r$ with $a = 0.6$, $b = -0.6$, $\mathfrak{c}_1 = 7.8$ and $\gamma = 0.3$.}
\label{feos2f2}
\end{figure}

\begin{figure} 
\centering
\includegraphics[scale=0.65]{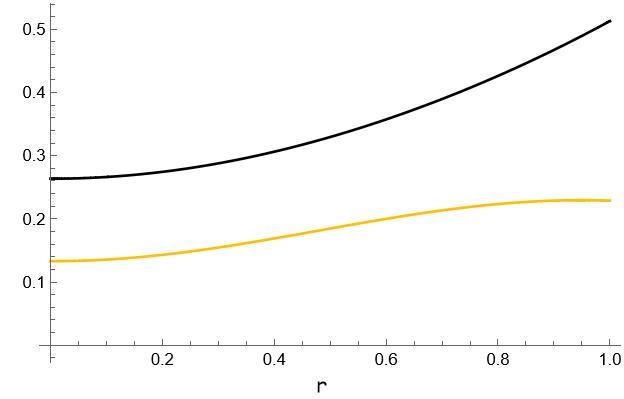}
\caption{$dp_r/d\mu$ (black) and $dp_{\perp}/d\mu$ (yellow) for fluid 2 for the Florides-class solution in \eqref{feos2a} for fluid 2 vs $r$ with $a = 0.6$, $b = -0.6$, $\mathfrak{c}_1 = 7.8$ and $\gamma = 0.3$.}
\label{feos2af}
\end{figure}

The energy densities of both fluids are 
\begin{subequations}\label{fe2ad}
\begin{align}
\mu_{1} =& \frac{1}{3} \gamma  c_1 \left(12 a^2+3 a (1-4 b) r^2+2 b (2 b-1) r^4\right)^{\mathfrak{z}_3} \exp \beta,\\
\mu_{2} = & \frac{1}{12} \left(\frac{(4 b+1) \left(9 a^2-7 a b r^2+2 b^2 r^4\right)}{\left(b r^2-a\right)^3} \right.\nonumber\\
& \left. -4 \gamma  c_1 \left(12 a^2+3 a (1-4 b) r^2+2 b (2 b-1) r^4\right)^{\mathfrak{z}_3} \exp \beta \right),
\end{align}
\end{subequations}
where we have used \eqref{fz3} and \eqref{beta}. The energy densities are shown in Fig. \ref{fe1d}. The pressures for both fluids are plotted in Figs. \ref{feos2f1}, \ref{feos2f2} and \ref{feos2af}. The radial pressure of fluid 2 approaches zero before fluid 1. This solution represents a shell.
%%%%%%%%%%%%%%%%%%%%%%%%%%%%%%%%%%%%%%%%%%%%%%%%%%%%%%%%%%%%%%%%%%%%%%%%%%%%%%%%%%%%%%%%%%%%%%
%%%%%%%%%%%%%%%%%%%%%%%%%%%%%%%%%%%%%%%%%%%%%%%%%%%%%%%%%%%%%%%%%%%%%%%%%%%%%%%%%%%%%%%%%%%%%%
%%%%%%%%%%%%%%%%%%%%%%%%%%%%%%%%%%%%%%%%%%%%%%%%%%%%%%%%%%%%%%%%%%%%%%%%%%%%%%%%%%%%%%%%%%%%%%
%%%%%%%%%%%%%%%%%%%%%%%%%%%%%%%%%%%%%%%%%%%%%%%%%%%%%%%%%%%%%%%%%%%%%%%%%%%%%%
\begin{figure} 
\centering
\includegraphics[scale=0.65]{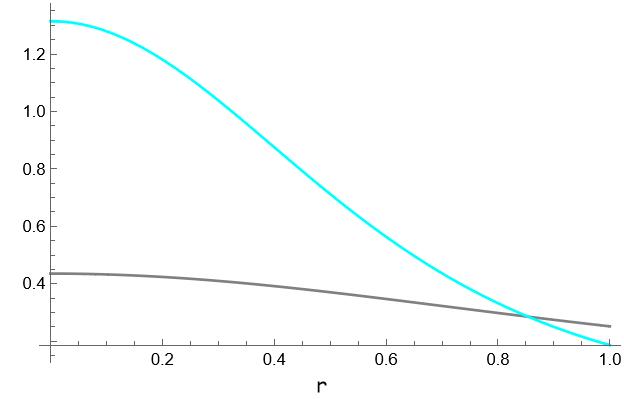}
\caption{Energy density profile for the Florides-class solution in \eqref{fe2ad} for fluid 1 (gray) and fluid 2 (cyan) vs $r$ with $a = 0.6$, $b = -0.6$, $\mathfrak{c}_1 = 7.8$ and $\gamma = 0.3$.}
\label{fe2d}
\end{figure}
%%%%%%%%%%%%%%%%%%%%%%%%%%%%%%%%%%%%%%%%%%%
\section{Solution Reconstruction}\label{SR}
%%%%%%%%%%%%%%%%%%%%%%%%%%%%%%%%%%%%%%%%%%%
In \cite{ncd1}, we formulated a reconstruction algorithm for the two-fluid isotropic TOV equations, by expanding the technique that was proposed in \cite{sante1,sante2} to obtain new two-fluid solutions starting from a given metric. In this section, we generalize this technique to include anisotropy.

We start by rearranging (\ref{DE2}) as
\begin{equation}
X = - \frac{1}{2} - \frac{\mathcal{K}_{, \rho}}{2 \mathcal{K}}. \label{X}
\end{equation}
Equation (\ref{c1}) gives the relation
\begin{equation}
\mathbb{M}_{tot} = -\frac{1}{4} + \mathcal{K} + \frac{\mathcal{K}_{,\rho}}{2 \mathcal{K}},\label{92}
\end{equation}
Utilizing (\ref{DE1}) together with (\ref{92}), we obtain
\begin{equation}
\begin{aligned}
{P}_{tot}&=P_1 + P_2 \\
&= \frac{1}{3} \left[2 Y_{, \rho} + 2 Y^2 + Y \right] - \frac{1}{3} \mathcal{K} - \left(\frac{2 Y + 1}{6} \right)\frac{\mathcal{K}_{, \rho}}{\mathcal{K}}+ \frac{1}{12}. \label{93}
\end{aligned}
\end{equation}
The total anisotropic pressure is obtained from (\ref{c2}) with \eqref{93}, and is given by
\begin{equation}\label{addcon}
\mathbb{P}_{tot} = \frac{\mathcal{K}_{, \rho}-4 \mathcal{K}^2+\mathcal{K}-4 \mathcal{K} Y+2 Y \mathcal{K}_{, \rho}-4 \mathcal{K} Y^2+4 \mathcal{K} Y}{6 \mathcal{K}}.
\end{equation}
In the isotropic case the condition $\mathbb{P}_{tot}=0$ sets the first order relationship 
\begin{equation}\label{iso_constr}
0= \frac{\mathcal{K}_{, \rho}-4 \mathcal{K}^2+\mathcal{K}-4 \mathcal{K} Y_{, \rho}+2 Y \mathcal{K}_{, \rho}-4 \mathcal{K} Y^2+4 \mathcal{K} Y}{6 \mathcal{K}}.
\end{equation}
between $\mathcal{K}$ and $Y$. Hence, whenever this relationship is violated, the metric will be only compatible with anisotropic fluids. In seeking physically viable solutions, it is convenient to violate the constraint \eqref{iso_constr} using $Y$ and $\mathcal{K}$ from known physically viable isotropic solutions. This will increase the probability of obtaining a well-behaved reconstructed solution.    

The analysis in \cite{ncd1} in the case of two isotropic fluids showed a degeneracy in the equations above as a given spacetime metric might correspond to many different combinations of fluids. However, as the conservation equations for the individual fluids are independent equations, one can use these relations to break this degeneracy. The same happens in the anisotropic case. Thus we need to add either \eqref{fl1_cons} or \eqref{fl2_cons} to our reconstruction equations.  Without loss of generality, we can choose \eqref{fl1_cons}  i.e. the equation for fluid 1.  We then obtain, setting the interaction to zero,
\begin{equation}
\begin{split} \label{psr}
{P}_{1, \rho} + \mathbb{P}_{1, \rho} = &- Y (\mathbb{M}_1 + P_1) - \mathbb{P}_1 \left( 2X + Y + \frac{3}{2}\right) \\
& - 2 X P_1,
\end{split}
\end{equation}
a first order differential equation. 

Depending on the properties of the source fluids, there are different possible resolution approaches. 
For example, in cases where one fluid is isotropic, we can obtain $\mathbb{P}_{tot}=\mathbb{P}_{2}$ from \eqref{addcon}, if the metric and, therefore, $Y$ and $\mathcal{K}$ are assigned. At this point, we can derive $P_{tot}$ using \eqref{93}. Then $P_1$ can be determined with an ansatz such as the equation of state, while $P_2$ is derived by subtracting $P_1$ from $P_{tot}$. In addition, $\mathbb{M}_{tot}$ is derived from the constraint \eqref{c2} with \eqref{92}. 

For the cases in which both fluids are anisotropic, additional information is required. This can be obtained, for example, by assuming an equation of state relating the anisotropic pressure $\mathbb{P}$ to the density $\mathbb{M}$ and pressure $P$, or by giving a relation between $\mathbb{M}$ and $\mathcal{K}$. 

Despite the sheer number of potential combinations of solutions one could use in the reconstruction, there is no guarantee that the solution obtained will be physically viable. Here, we will limit ourselves to illustrate one example. In particular, we will perform the reconstruction using the interior Schwarzschild and Tolman IV solutions. 

We choose $Y$ from the Interior Schwarzschild solution and $\mathcal{K}$ from the Tolman IV solution:
\begin{subequations}
\begin{align}
Y &= \frac{e^{\rho} \mu_1}{-2 c_1 \sqrt{3-e^{\rho} \mu_1}+2 e^{\rho} \mu_1-6}, \\
\mathcal{K} &= -\frac{R^2 \left(A^2+2 e^{\rho} \right)}{4 \left(A^2+e^{\rho} \right) \left(e^{\rho} -R^2\right)},
\end{align}
\end{subequations}
and we assume the equation of state
\begin{equation} \label{eosiso}
\mathbb{M}_1 = a P_1,
\end{equation}
where $a$ is a constant. We assume fluid 1 to be isotropic, so $\mathbb{P}_1 = 0$. Eq. \eqref{psr} reduces to
\begin{equation} \label{psriso}
{P}_{1, \rho} = - Y (\mathbb{M}_1 + P_1) - 2 X P_1.
\end{equation}
As mentioned above with $Y$ and $\mathcal{K}$ we can determine all the total quantities.  Then, using \eqref{eosiso} we can solve \eqref{psriso}  for $P_{1}$. In addition, $P_2$ can be obtained by subtracting $P_1$ from $P_{tot}$ and, since $\mathbb{P}_1 = 0$,  $\mathbb{P}_2 = \mathbb{P}_{tot}$. Finally, $\mathbb{M}_2$ is obtained by subtracting $\mathbb{M}_1$ from $\mathbb{M}_{tot}$. For the reconstructed quantities, we have \eqref{RTS0} in Appendix \ref{AASR}.

\begin{figure} 
\centering
\includegraphics[scale=0.55]{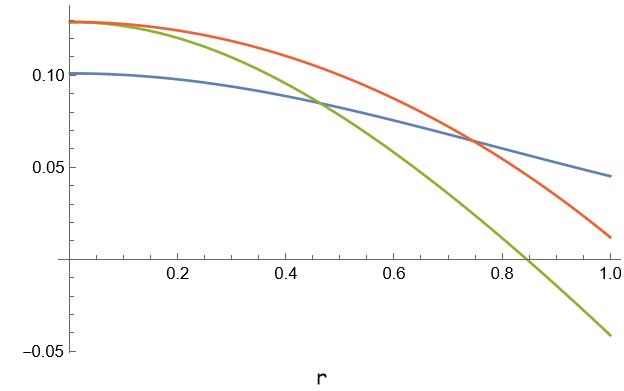}
\caption{Radial pressure for the solution in \eqref{ST0} for fluid 1 (blue) and fluid 2 (green), and tangential pressure for fluid 2 (red) vs $r$ with $a = 3$, $\mu_1 = 0.9$, $R = 1.75$, $A = 1.95$, $\mathfrak{c}_1 = -0.2$, and $c_1 = -3$.}
\label{f10a}
\end{figure}

\begin{figure} 
\centering
\includegraphics[scale=0.55]{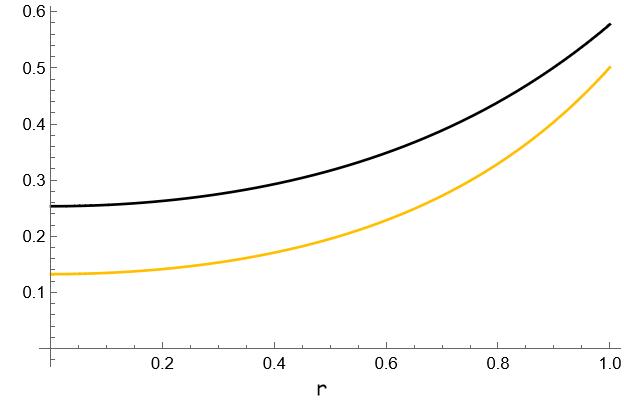}
\caption{$dp_r/d\mu$ (black) and $dp_{\perp}/d\mu$ (yellow) for the solution in \eqref{ST0} for fluid 2 vs $r$ with $a = 3$, $\mu_1 = 0.9$, $R = 1.75$, $A = 1.95$, $\mathfrak{c}_1 = -0.2$, and $c_1 = -3$.}
\label{f10b}
\end{figure}

To obtain $p_1, p_2, \Pi_1,$ and $\Pi_2$ in terms of $r$, we use $\phi$ for the Tolman IV solution \eqref{phit}, with \eqref{newvars}. Finally, using \eqref{prpt}, we can write the radial and tangential pressures for each fluid in terms of $r$.  These quantities are too long to be reported here, and are given in \eqref{ST0} of Appendix \ref{AASR}. The energy densities of both fluids are given in \eqref{f10ed} of Appendix \ref{AASR} and are shown in Fig. \ref{f10d}. The pressures for both fluids are plotted in Fig. \ref{f10a}, while Fig. \ref{f10b} shows the square of the sound speed for fluid 2 (we have omitted fluid 1 from the plot since this quantity has a constant value of $\frac{1}{a}$ for fluid 1). Since the radial pressures of both fluids vanish at different values of $r$ we can conclude that this solution represents a shell.
%%%%%%%%%%%%%%%%%%%%%%%%%%%%%%%%%%%%%%%%%%%%%%%%%%%%%%%%%%%%%%%%%%%%%%%%%%%%%%%
%%%%%%%%%%%%%%%%%%%%%%%%%%%%%%%%%%%%%%%%%%%%%%%%%%%%%%%%%%%%%%%%%%%%%%%%%%%%%%%
%%%%%%%%%%%%%%%%%%%%%%%%%%%%%%%%%%%%%%%%%%%%%%%%%%%%%%%%%%%%%%%%%%%%%%%%%%%%%%
\begin{figure} 
\centering
\includegraphics[scale=0.65]{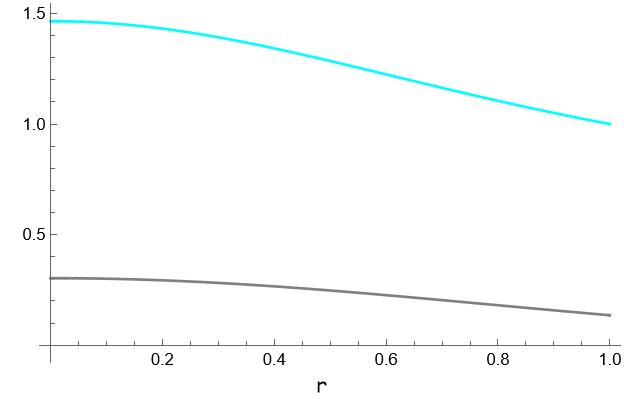}
\caption{Energy density profile for the solution in \eqref{f10ed} for fluid 1 (gray) and fluid 2 (cyan) vs $r$ with $a = 3$, $\mu_1 = 0.9$, $R = 1.75$, $A = 1.95$, $\mathfrak{c}_1 = -0.2$, and $c_1 = -3$.}
\label{f10d}
\end{figure}
%%%%%%%%%%%%%%%%%%%%%%%%%%%%%%%%%%%%%%%%%%%%%%%%%%%%%%%%%%%%%%%%%%%%%%%%%%
\section{Generating theorems}\label{GT}
%%%%%%%%%%%%%%%%%%%%%%%%%%%%%%%%%%%%%%%%%
Boonserm {\it et al}. developed generating/transformation theorems which ``map one perfect fluid sphere into another'' \cite{boon1, boon2}. The term perfect fluid sphere represents a static, isotropic, spherically symmetric fluid distribution in curved spacetimes which, in turn, can be associated to the relativistic stellar objects described by the TOV equations. Using generating theorems, one can start from a known solution of the TOV equations and obtain new solutions. Carloni {\it et al}. demonstrated that in the context of covariant TOV equations, the generating theorems are equivalent to a linear deformation of a given solution \cite{sante1,sante2}. In this section, we will show that these theorems also exist in the case of anisotropic multifluid solutions. 

Indeed one can find several different generating theorems, depending on the quantities that can be ``deformed''. 
For brevity, we will show explicitly some specific cases, leaving a complete taxonomy for a future work. The selected cases will illustrate: (i) deforming a two-fluid anisotropic solution to a new two-fluid anisotropic solution, (ii) deforming an isotropic solution to obtain an anisotropic one, and finally, (iii) starting with two isotropic fluids and obtaining a solution with more than two anisotropic fluids.

For all of these examples, we will be deforming the quantity $Y$ as 
\begin{equation}\label{ydeform}
Y \rightarrow \bar{Y} + \tilde{Y},
\end{equation}
where, here and in the following, the bar indicates the known solution and tilde represents the deformed quantities. We can then use the definition of $Y$ in terms of the metric coefficient $k_1$ in \eqref{vars}, to obtain the  transformation of the metric that corresponds to this deformation:
\begin{subequations}\label{gtmetric}
\begin{align}
k_1 &\rightarrow a_1 \bar{k}_1  \exp \left(2 \int{\tilde{Y}}d\rho \right),\\
k_2 &\rightarrow \bar{k}_2,\\
k_3 &\rightarrow \bar{k}_3,
\end{align}
\end{subequations}
where $a_{1}$ is a constant of integration. 

In what follows,  we will connect the deformation of $Y$ given by \eqref{ydeform} to deformations in other geometrical and thermodynamical quantities. We will even consider a case in which the deformation allows the mapping to a solution with an additional fluid. The deformed matter variables, together with the new metric coefficients, represent the new solution.

%%%%%%%%%%%%%%%%%%%%%%%%%%%%%%%%%%%%%%%%%%%%%%%%%%%%%%%%%%%%%%%%%%
\subsection{Case 1: Two anisotropic fluids $\rightarrow$ Two anisotropic fluids} \label{GT1}
%%%%%%%%%%%%%%%%%%%%%%%%%%%%%%%%%%%%%%%%%%%%%%%%%%%%%%%%%%%%%%%%%%%%%%%%%%%%%%%%%%%%%%%%%%%%%%
As a first example we will start from a solution of the TOV equations consisting of two anisotropic fluids.

The $Y$ deformation \eqref{ydeform} gives a new solution of Eqs. \eqref{tov2} if the variables in the TOV equations are deformed in the following way

\begin{subequations}\label{gt1def}
\begin{align}
{P}_1 &\rightarrow \bar{{P}}_{1} ,\\
{P}_2 &\rightarrow \bar{{P}}_{2},\\
\mathbb{P}_1 &\rightarrow \bar{\mathbb{P}}_{1} + \tilde{\mathbb{P}}_{1},\\
\mathbb{P}_2 &\rightarrow \bar{\mathbb{P}}_{2},\\
\mathbb{M}_1 &\rightarrow \bar{\mathbb{M}}_{1} + \tilde{\mathbb{M}}_{1},\\
\mathbb{M}_2 &\rightarrow \bar{\mathbb{M}}_{2} - \tilde{\mathbb{M}}_{1},\\
{\mathcal{K}}&\rightarrow\bar{\mathcal{K}}.
\end{align}
\end{subequations}
Notice that $\mathbb{M}_{tot}$ is not deformed. Let us now prove that the deformations can be obtained in an analytical, albeit formal way. Substituting the deformations in the constraint \eqref{c2} gives
\begin{equation}\label{gt1py}
\tilde{Y} = \tilde{\mathbb{P}}_{1},
\end{equation}
i.e. the deformation in $Y$ has the same analytical form as the deformation in the anisotropic pressure of fluid 1. Using this result, and taking into account that Eqs. \eqref{tov2} hold for the barred quantities, the deformed Eq. \eqref{EqPtot} takes the form
\begin{eqnarray}\label{gt1pb}
\tilde{\mathbb{P}}_{1, \rho} &=& -\frac{1}{4} \tilde{\mathbb{P}}_{1} \left(12 \bar{\mathcal{K}}-4 \bar{\mathbb{M}}_{1}-4 \bar{\mathbb{M}}_{2}+8 \bar{P}_{1}+8 \bar{\mathbb{P}}_{1} \right. \nonumber \\
&&\left. +8 \bar{P}_{2}+8 \bar{\mathbb{P}}_{2}-1 \right)-\tilde{\mathbb{P}}_{1}^2,
\end{eqnarray}
which is a Bernoulli equation for $\tilde{\mathbb{P}}_{1}$ and has a formal general solution.

We now need to determine the energy densities of the fluids in the new solution. Substituting the deformations in the TOV equation for fluid one \eqref{p1} and remembering that the same equation holds for the barred variables, we can solve for  $\tilde{\mathbb{M}}_{1}$ to obtain
\begin{eqnarray}\label{gt1m1d}
\tilde{\mathbb{M}}_{1} &=&\frac{\tilde{\mathbb{P}}_{1} }{4 \left(\tilde{\mathbb{P}}_{1}+\bar{Y} \right)} \left(12 \bar{\mathcal{K}}+4 \bar{\mathbb{M}}_{2}+12 \bar{P}_{1} +12 \bar{\mathbb{P}}_{1}+16 \bar{P}_{2} \right. \nonumber \\
&& \left. +16 \bar{\mathbb{P}}_{2}-12 \bar{Y} -3 \right).
\end{eqnarray}
Hence with the above procedure, we have started with two anisotropic fluids and, after the deformation, we have a modified geometry and two new anisotropic fluids. Both fluids have energy densities that are different to those of the known solution and one of them has a modified anisotropic pressure. 

Notice that in performing the above procedure, we need to ensure that all thermodynamical variables are positive, so we must impose that $\mathbb{M}_{2}$ is positive as an additional condition on the solution of the above equations. This is achieved by verifying that:

\begin{equation}
\bar{\mathbb{M}}_{2} - \tilde{\mathbb{M}}_{1} >0.
\end{equation}
%%%%%%%%%%%%%%%%%%%%%%%%%%%%%%%%%%%%%%%%%%%%%%%%%%%%%%%%%%%%%%%%%%%%%%%%
\subsection{Case 2: Two isotropic fluids $\rightarrow$ Two anisotropic fluids}\label{GT2}
%%%%%%%%%%%%%%%%%%%%%%%%%%%%%%%%%%%%%%%%%%%%%%%%%%%%%%%%%%%%%%%%%%%%%%%%%%%%%%%%%%%%%%%%%%%%%%
As a second example, we consider a solution comprising of two isotropic fluids ($\bar{\mathbb{P}}_1 = 0$ and $\bar{\mathbb{P}}_2 = 0$). We now deform $Y$ as in \eqref{ydeform} and we require that after the deformation the new solution will have two  {\it anisotropic} fluids with the same energy density. This result can be achieved by assuming that the fluid and geometry variables are deformed as
\begin{subequations}\label{gt2def}
\begin{align}
{P}_1 &\rightarrow \bar{{P}}_{1} ,\\
{P}_2 &\rightarrow \bar{{P}}_{2},\\
\mathbb{P}_1 &\rightarrow \tilde{\mathbb{P}}_{1},\\
\mathbb{P}_2 &\rightarrow \tilde{\mathbb{P}}_{2},\\
\mathbb{M}_1 &\rightarrow \bar{\mathbb{M}}_{1},\\
\mathbb{M}_2 &\rightarrow \bar{\mathbb{M}}_{2},\\
{\mathcal{K}}&\rightarrow\bar{\mathcal{K}}.
\end{align}
\end{subequations}

Applying these deformations and taking into account that \eqref{c2} holds for the barred quantities associated to the starting solution, we obtain
\begin{equation}\label{gt2py}
\tilde{Y} = \tilde{\mathbb{P}}_{1} + \tilde{\mathbb{P}}_{2} = \tilde{\mathbb{P}}_{tot}.
\end{equation}

At this point, since the barred quantities satisfy by themselves the TOV equations, the equation for the total pressure \eqref{EqPtot} can be written as a Bernoulli equation for the total deformed anisotropic pressure $\tilde{\mathbb{P}}_{tot}$

\begin{eqnarray}
&&\tilde{\mathbb{P}}_{tot} (12 \bar{\mathcal{K}} - 4 \bar{\mathbb{M}}_{1} - 4 \bar{\mathbb{M}}_{2} + 8 \bar{P}_{1} + 8 \bar{P}_{2}-1) \nonumber \\
&&+4 \tilde{\mathbb{P}}_{tot, \rho}+4 \tilde{\mathbb{P}}_{tot}^2=0.
\end{eqnarray}

The TOV equation for fluid 1, \eqref{p1}, under deformation gives the following equation for $\tilde{\mathbb{P}}_{1}$
\begin{eqnarray}\label{gt2pb}
\tilde{\mathbb{P}}_{1, \rho} &=& \tilde{\mathbb{P}}_{1} \left(2 \bar{\mathbb{M}}_{1}+2 \bar{\mathbb{M}}_{2} +2 \bar{P}_{1} +2 \bar{P}_{2}-\tilde{\mathbb{P}}_{tot} \right. \nonumber\\
&& \left. -3 \bar{Y}-\frac{1}{2}\right) -\tilde{\mathbb{P}}_{tot} (\bar{\mathbb{M}}_{1}+3 \bar{P}_{1}).
\end{eqnarray}

Then we can derive $\tilde{\mathbb{P}}_{2}$ using Eq. \eqref{gt2py}. As the Bernoulli equations have formal general solutions we can conclude that we have obtained a new solution with two anisotropic fluids.
%%%%%%%%%%%%%%%%%%%%%%%%%%%%%%%%%%%%%%%%%%%%%%%%%%%%%%%%%%%%%%%%%%%%%%%%%%%%
\subsection{Case 3: Two isotropic fluids $\rightarrow$ Three anisotropic fluids}\label{GT3}
%%%%%%%%%%%%%%%%%%%%%%%%%%%%%%%%%%%%%%%%%%%%%%%%%%%%%%%%%%%%%%%%%%%%%%%%%%%%%%
Our final example involves a change in both the nature and number of fluids. We begin with two isotropic fluids and we perform the deformation 
\begin{subequations}\label{gt3def}
\begin{align}
{P}_1 &\rightarrow \bar{P}_{1} + {P}_{3},\\
{P}_2 &\rightarrow \bar{P}_{2},\\
\mathbb{P}_1 &\rightarrow \mathbb{P}_{3},\\
\mathbb{P}_2 &\rightarrow -\mathbb{P}_{3},\\
\mathbb{M}_1 &\rightarrow \bar{\mathbb{M}}_{1},\\
\mathbb{M}_2 &\rightarrow \bar{\mathbb{M}}_{2},\\
{\mathcal{K}}&\rightarrow\bar{\mathcal{K}}.
\end{align}
\end{subequations}
where now, we consider $P_3$ not as a change in $P_1$, but as the pressure of an {\it additional} fluid present only in the new solution. In this picture, the pressure $P_3$ must satisfy an equation similar to Eq. \eqref{p1} for $P_1$  and  Eq. \eqref{p2} for $P_2$:
\begin{equation}\label{gt3sys}
P_{3, \rho}+ \mathbb{P}_{3, \rho} = P_3 (-2 X-Y)-\mathbb{P}_3 \left(2 X+Y+\frac{3}{2}\right)-\mathbb{M}_3 Y.
\end{equation}
We will assume: (i) that the third fluid is anisotropic, so in the above equation $\mathbb{P}_3\neq 0$, and (ii) after the deformation, fluids 1 and 2 are anisotropic as in \eqref{gt3def}. Notice that as we have seen in Sec. \ref{CON}, the anisotropic term $\Pi$ can be negative, so we can consider $\mathbb{P}_1$ or $\mathbb{P}_2$ less than zero.

Applying these deformations and solving the constraint \eqref{c2} with the initial solution, we obtain the relation
\begin{equation}\label{gt3py}
\tilde{Y} = P_3.
\end{equation}
Applying the deformation and the result \eqref{gt3py} to Eq. \eqref{p1} with no interaction and remembering that the barred quantities satisfy \eqref{p1}, we obtain the following equation for $P_3$
\begin{equation}\label{gt3p3}
\begin{split}
P_{3, \rho}  =- &P_3^2 - \bigg(3 \bar{\mathcal{K}} - \bar{\mathbb{M}}_{1} - \bar{\mathbb{M}}_{2}+2 \bar{P}_{1}+2 \bar{P}_{2} -\frac{7}{4}\bigg),\end{split}
\end{equation}
which is a Bernoulli equation. Now, substituting the above results and Eq. \eqref{constrX} into \eqref{gt3sys}, we obtain the equation 
\begin{eqnarray}\label{gt3pb3}
\mathbb{P}_{3, \rho} &=& \mathbb{P}_3 \left[\frac{1}{4} \left(8 \bar{\mathbb{M}}_{1}+8 \bar{\mathbb{M}}_{2}+8 \bar{P}_{1}+8 \bar{P}_{2}-12 \bar{Y}-2 \right) \right.\nonumber \\
&& \left. -P_3\right] +\frac{1}{4} P_3 \left(12 \bar{\mathcal{K}} + 4 \bar{\mathbb{M}}_{2}+12 \bar{P}_{1}+16 \bar{P}_{2} \right. \nonumber \\
&& \left. -12 \bar{Y} -3\right).
\end{eqnarray}
Finally, substituting \eqref{gt3p3} and \eqref{gt3pb3} into  \eqref{EqPtot}
we can solve for $\mathbb{M}_{3}$:
\begin{equation}\label{gt3m3}
\mathbb{M}_{3} = \frac{P_3(\bar{\mathbb{M}}_{1}+\bar{P}_{1})}{P_3+\bar{Y}}.
\end{equation}
Hence, we have obtained three anisotropic fluids from the deformation of two isotropic fluids. The pressure $P$ of the first fluid was deformed such that it contained the pressure of the known solution and an additional term which represents the pressure of the third fluid. The pressure of the second fluid was not deformed, and the anisotropic pressures of both fluid 1 and 2 were deformed such that they contained the anisotropic pressure of fluid 3. The energy densities of fluids 1 and 2 were not deformed, and the energy density of fluid 3 was calculated.
%%%%%%%%%%%%%%%%%%%%%%%%%%%%%%%%%%%%%%%%%%%%%%%%%%%%%%%%%%%%%%%%%%%%%%%%%%%%
\section{Discussion and conclusion}\label{CR}
%%%%%%%%%%%%%%%%%%%%%%%%%%%%%%%%%%%%%%%%%%%%%%%%%%%%%%%%%%%%%%%%%%%%%%%%%%%%
In this paper, we have presented the 1+1+2 semi-tetrad equations for the two-fluid anisotropic case, along with its generalisation to $N$-fluids, and we have used these equations to construct a covariant multifluid version of the TOV equations. Such generalization is needed because anisotropy is a physical property encountered in the study of most compact objects.

The general structure of these equations revealed several interesting facts. For example, the weak energy condition showed explicitly that the anisotropic pressure term $\Pi$ can be negative, provided that \eqref{ecnew} holds. Allowing for negative $\Pi$ implies that the the non radial pressure in an object is bigger than the radial one. In addition, at the boundary, the junction conditions require that the radial pressure vanishes, but there is no restriction on the behaviour of the tangential pressure. Therefore, the tangential pressure does not necessarily need to vanish at the boundary.

In order to solve the covariant TOV equations analytically, we have proposed several resolution strategies in the case of two fluids. In particular, using some well-known single fluid solutions we explored some direct resolution algorithms as well as some indirect ones, namely, reconstruction methods and generating theorems.

As a first example we derived the two-fluid generalization of the simplest known single fluid anisotropic solution: the Bowers-Liang solution. Our generalization has the same geometry ad the original solution, but is composed of two different fluids. Its structure allows one to see explicitly the changes induced on the thermodynamical quantities by the presence of anisotropy. Specifically, the radial pressure becomes lower at the core and higher on the surface, when compared to the isotropic case. This change in the radial pressure implies that the matter composition at the core is different, i.e., there is a different equation of state for this interior region.  This gives the chance to observe explicitly the structural changes \cite{viaggiu} that anisotropy induces in a given object.

We then applied the same strategy to other known solutions. From an element of the Florides-class of solutions, we were able to obtain six new anisotropic two-fluid solutions. In some of these solutions, one of the fluids was barotropic, a feature which does not seem to be possible in the single fluid case or when all fluids are isotropic. None of these solutions, however, could represent an entire object as they could not be smoothly soldered to a vacuum exterior. As we have commented, solutions such as these represent only a specific shell in the compact object and requires matching to an additional shell before being connected to an exterior solution, such as the Schwarzschild \cite{1916skpa.conf.424S} or Vaidya \cite{vaidya} solution. Such ``shell structure'' is expected to arise in multifluid systems, in which the properties of matter induces layers with different mechanical and thermodynamical properties.

A less straightforward approach, which yields interesting results, is given by the so-called reconstruction techniques. We found that it is possible to extend the reconstruction method proposed in \cite{sante1,sante2} to the case of anisotropic multifluid objects. In particular, we found that by mixing the metric components of different well-known isotropic solutions one can reconstruct new multifluid anisotropic solutions with desirable features. We presented one solution of this type, constructed from the interior Schwarzschild and Tolman IV solutions.

Finally, we considered the extension of the generating theorems to the case of multifluid anisotropic solutions. It turns out that one can find a great number of such theorems that are able to connect solutions with different types of fluids or even with a different number of fluids. We have limited ourselves here to three specific examples. A first, in which a solution comprising two anisotropic fluids is mapped to a new solution in which one of the fluids has a different anisotropic pressure. A second, in which a two-fluid anisotropic solution is obtained from a two-fluid isotropic one. Finally a third example, in which these theorems are used to change the number and character of fluids in a given solution. As it is well known, the generating theorems can be used to form chains of solutions of the TOV equations. We find here that this is even more true in the case of multifluid anisotropic solutions. We will extend this investigation and examine some of the other variations to these theorems in a future work. 

%%%%%%%%%%%%%%%%%%%%%%%%%%%%%%%%%%%%%%%%%%%%%%%%%%%%%%%%%%%%%%%%%%%%%%%%%
\begin{acknowledgments}
NFN would like to acknowledge funding from the CoE-MaSS and Oppenheimer Memorial Trust.
\end{acknowledgments}

\appendix
%%%%%%%%%%%%%%%%%%%%%%%%%%%%%%%%%%%%%%%%%%%%%%%%%%%%%%%%%%%%%%%%%%%%%%%%%%%%
\section{Additional details of the 1+1+2 covariant approach for a single fluid}\label{AA}
%%%%%%%%%%%%%%%%%%%%%%%%%%%%%%%%%%%%%%%%%%%%%%%%%%%%%%%%%%%%%%%%%%%%%%%%%%%%
In this section we present the main aspects of the 1+1+2 covariant approach that are not covered in Sec. \ref{TFC}. The reader is referred to \cite{clarkbar,clarkson,betschart} for more details. 

The complete list of kinematical variables is given by 
\begin{subequations} 
\begin{align} 
&& \mathcal{A} = e_a \dot{u}^a \hspace{0.2cm},\hspace{0.2cm} \mathcal{A}_a = N_{ab} \dot{u}^b \hspace{0.2cm},\hspace{0.2cm} \Theta = D_a u^a ,   \\
&& \Omega = \frac{1}{2} \varepsilon^{abc} D_{[a} u_{b]} e_c \hspace{0.2cm},\hspace{0.2cm}{\Omega}^a = \frac{1}{2} \epsilon^{cbd} D_{[c} u_{b]} {N_d}^a,   \\
&& \Sigma = {\sigma}^{ab} e_a e_b  \hspace{0.2cm},\hspace{0.2cm} {\Sigma}_a = {\sigma}_{cd} e^d {N_a}^c,\\
&& {\Sigma}_{ab} = \left({N^c}_{(a}{N_{b)}}^d - \frac{1}{2} N_{ab} N^{cd} \right){\sigma}_{cd},   \\
&& {\sigma}_{ab} = \left({h^c}_{(a} {h_{b)}}^d - \frac{1}{3} h_{ab} h^{cd} \right)D_c u_d,   \\
&& a_b = e^c D_c e_b = \hat{e}_b \hspace{0.2cm},\hspace{0.2cm} \phi = {\delta}_a e^a,   \\
&& {\zeta}_{ab} = \left({N^c}_{(a}{N_{b)}}^d - \frac{1}{2} N_{ab} N^{cd} \right){\delta}_c e_d,   \\
&& \xi = \frac{1}{2} \varepsilon^{ab} {\delta}_a e_b,   \\
&& \mathcal{E} = {C}_{acbd} u^c u^d e^a e^b,   \\
&& \mathcal{E}_a = C_{cdef} u^e u^f e^c {N^d}_a \hspace{0.2cm},\hspace{0.2cm} \mathcal{E}_{ab} = C_{\{acb\}d} u^c u^d,   \\
&& \mathcal{H} = \frac{1}{2} {\varepsilon}_{ade} {C^{de}}_{bc} u^c e^a e^b,   \\
&& \mathcal{H}_a = \frac{1}{2}\varepsilon_{cbe} {C^{be}}_{df} u^f e^c {N^d}_a,   \\
&& \mathcal{H}_{ab} = \frac{1}{2}{\varepsilon_{\{ade}} {C^{de}}_{b\}f} u^f,
\end{align}
\end{subequations} 
where $\varepsilon_{ab} \equiv \varepsilon_{abc} e^c$ and $\varepsilon_{abc} = \eta_{dabc} u^d$ are the volumes of the two hypersurfaces, and $C_{cdef}$ is the Weyl tensor. The symmetrisation over the indices of a tensor is represented as $T_{(ab)} = \frac{1}{2} (T_{ab} + T_{ba})$, and the anti-symmetrisation as $T_{[ab]} = \frac{1}{2} (T_{ab} - T_{ba})$. We use curly brackets $\{ \}$ to denote the Projected Symmetric Trace-Free part of a tensor with respect to $e^a$:
\begin{equation}
X_{\{ab\}} 
\equiv \left( N^{c}{}_{(a}N_{b)}{}^{d} - \frac{1}{2}N_{ab} N^{cd}\right) \mathds{X}_{cd}.
\end{equation} 
According to \cite{betschart}, the remaining 1+1+2 scalars which fully describe the spacetime can be divided into 3 categories:
\begin{subequations} \label{hatsystem1}
\begin{align} 
\intertext{\it Propagation:} \nonumber \\
\hat{\phi} &= -\frac{1}{2} {\phi}^2 - \frac{2}{3} \mu - \mathcal{E} - \frac{1}{2} \Pi, \\
Q &= 0, \\ 
\hat{\mathcal{E}} - \frac{1}{3} \hat{\mu} + \frac{1}{2} \hat{\Pi} &= -\frac{3}{2} {\phi} \left(\mathcal{E} + \frac{1}{2} \Pi \right). \\ 
\intertext{\it Evolution:} \nonumber \\
0 &= -\mathcal{A} \phi + \frac{1}{3}(\mu + 3p)- \mathcal{E} + \frac{1}{2} \Pi, \\
0 &= \frac{1}{2} \phi Q. \\
\intertext{\it Propagation/evolution:} \nonumber \\
\hat{\mathcal{A}} &= - \mathcal{A} \left( \mathcal{A} + \phi \right) + \frac{1}{2}(\mu + 3p), \\
\hat{Q} &= -Q \left(\phi + 2 \mathcal{A} \right) + j_u, \\
\hat{p} + \hat{\Pi} &= -\Pi \left(\frac{3}{2} \phi + \mathcal{A} \right) - \mathcal{A} \left( \mu + p \right) + j_e, \\
K &= \frac{1}{3} \mu - \mathcal{E} - \frac{1}{2} \Pi + \frac{1}{4} {\phi}^2, \\ \nonumber
\end{align}
\end{subequations}
where $\mu$, $p$, $Q$, and $\pi$ represent, in general, the total energy density, pressure, total heat flux, and total anisotropic pressure of the fluid respectively. In addition, we have included the total particle interaction currents as $j_u$ and $j_e$ according to the definition
\begin{equation} \label{j}
j_{a} = j_u u_a + j_e e_a,
\end{equation}
with $j_u$ and $j_e$ representing the $u_a$ and $e_a$ components, respectively.

Finally, combining equations \eqref{hatsystem1}, one can show that the Gauss curvature $K$ satisfies the propagation equation
\begin{equation} \label{Keq}
\hat{K}=-\phi K.
\end{equation}
%%%%%%%%%%%%%%%%%%%%%%%%%%%%%%%%%%%%%%%%%%%%%%%%%%%%%%%%%%%%%%%%%%%%
\begin{widetext}
\section{Multifluid 1+1+2 equations}\label{App2}
%%%%%%%%%%%%%%%%%%%%%%%%%%%%%%%%%%%%%%%%%%%%%%%%%%%%%%%%%%%%%%%%%%%%%%%%%%%%
In the case of a static spherically symmetric spacetime with $N$ different interacting fluids, the 1+1+2 equations are 
\begin{subequations} 
\begin{align} 
\hat{\phi} &=-\frac{1}{2} {\phi}^2 - \frac{2}{3} \sum_{i=1}^{N}{\mu}_i- \mathcal{E} - \frac{1}{2} \sum_{i=1}^{N}{\Pi}_i, \\
\sum_{i=1}^{N}Q_i &= 0, \\ 
\hat{\mathcal{E}} + \frac{3}{2} {\phi} \mathcal{E} &= \frac{1}{3}  \sum_{i=1}^{N}{\hat{\mu}}_i - \frac{1}{2} \sum_{i=1}^{N}{\hat{\Pi}}_i- \frac{3}{2} {\phi} \frac{1}{2} \sum_{i=1}^{N}{\Pi}_i, \\
\mathcal{E} + \mathcal{A} \phi&= \frac{1}{3}\sum_{i=1}^{N}({\mu}_i+3p_i) + \frac{1}{2} \sum_{i=1}^{N}{\Pi}_i,\\
K&=\frac{1}{4} {\phi}^2- \mathcal{E} +\frac{1}{3} \sum_{i=1}^{N}{\mu}_i- \frac{1}{2} \sum_{i=1}^{N}{\Pi}_i,\\
\hat{\mathcal{A}}&=- \mathcal{A} \left( \mathcal{A} + \phi \right) + \frac{1}{2} \sum_{i=1}^{N}({\mu}_i+3p_i), \\
 \hat{Q}_i&=-Q_i \left(\phi + 2 \mathcal{A} \right) +\sum_{k\neq i }^{N}j_u^{(i,k)},\label{Qi}\\
\hat{p}_i + \hat{\Pi}_i&=\sum_{k\neq i }^{N}j_e^{(i,k)} -{\Pi}_i \left(\frac{3}{2} \phi + \mathcal{A} \right) - \mathcal{A} \left( {\mu}_i + p_i\right),  \label{Pi}
\end{align}
\end{subequations}
where the index $i$ represents the $i$th component and $j_e^{(i,k)}$  is the interaction term between the component $i$ and $k$. Notice that $j_u^{(i,j)}=-j_u^{(j,i)}$ and $j_e^{(i,j)}=-j_e^{(j,i)}$.

Summing \eqref{Pi} over $i$ we have the equation for the total energy pressure: 
\begin{align}
\hat{p}_{tot}  + \hat{\Pi}_{tot}  &= -\sum_{i=1}^{N}\Pi_1 \left(\frac{3}{2} \phi + \mathcal{A} \right) - \mathcal{A} \sum_{i=1}^{N}\left( {\mu}_i + p_i \right).
\end{align}
In terms of the variable $\rho$, defined by the relation $K = K_0^{-1} e^{- \rho},$  the 1+1+2 equations become
\begin{subequations}
\begin{align} 
\phi {\phi}_{, \rho} &=-\frac{1}{2} {\phi}^2 - \frac{2}{3} \sum_{i=1}^{N}{\mu}_i- \mathcal{E} - \frac{1}{2} \sum_{i=1}^{N}{\Pi}_i, \\
\sum_{i=1}^{N}Q_i &= 0, \\
{\mathcal{E}}_{, \rho} + \frac{3}{2} \mathcal{E} &= \frac{1}{3}  \sum_{i=1}^{N}{{\mu}}_{i, \rho} - \frac{1}{2} \sum_{i=1}^{N}{{\Pi}}_{i, \rho} - \frac{3}{2} {\phi} \frac{1}{2} \sum_{i=1}^{N}{\Pi}_i, \\
\mathcal{E} + \mathcal{A} \phi&= \frac{1}{3}\sum_{i=1}^{N}({\mu}_i+3p_i) + \frac{1}{2} \sum_{i=1}^{N}{\Pi}_i,\\
K&=\frac{1}{4} {\phi}^2- \mathcal{E} +\frac{1}{3} \sum_{i=1}^{N}{\mu}_i- \frac{1}{2} \sum_{i=1}^{N}{\Pi}_i,\\
\phi{\mathcal{A}}_{, \rho}&=- \mathcal{A} \left( \mathcal{A} + \phi \right) + \frac{1}{2} \sum_{i=1}^{N}({\mu}_i+3p_i), \\
\phi \hat{Q}_{i, \rho}&=-Q_i \left(\phi + 2 \mathcal{A} \right) +\sum_{k\neq i }^{N}j_u^{(i,k)},\\
\hat{p}_i + \hat{\Pi}_i&=\sum_{k\neq i }^{N}j_e^{(i,k)} -{\Pi}_i \left(\frac{3}{2} \phi + \mathcal{A} \right) - \mathcal{A} \left( {\mu}_i + p_i\right).
\end{align}
\end{subequations}
Using the variables 
\begin{subequations}\label{newvarsb}
\begin{align}
X &= \frac{{\phi}_{, \rho}}{\phi}, & Y &= \frac{\mathcal{A}}{\phi}, & \mathcal{K} &= \frac{K}{{\phi}^2} ,\\
E &= \frac{\mathcal{E}}{{\phi}^2}, & \mathbb{M}_1 &= \frac{{\mu}_1}{{\phi}^2}, &
\mathbb{M}_2 &= \frac{{\mu}_2}{{\phi}^2}, \\
P_1 &= \frac{p_1}{{\phi}^2}, & P_2 &= \frac{p_2}{{\phi}^2}, &
\mathbb{P}_1 &= \frac{{\Pi}_1}{{\phi}^2} ,\\
\mathbb{P}_2 &= \frac{{\Pi}_2}{{\phi}^2} ,& \mathbb{Q}_1 &= \frac{Q_1}{{\phi}^2}, & 
\mathbb{Q}_2 &= \frac{Q_2}{{\phi}^2},
\end{align}
\end{subequations} 
we obtain the covariant TOV equations as
\begin{eqnarray}\label{tovN} 
{P}_{i, \rho} + \mathbb{P}_{i, \rho}   &=& \sum_{i\neq k}^N\mathbb{J}_e^{(i,k)}- {P_i}^2 - {\mathbb{P}_i}^2 + P_i \left[\mathbb{M}_i - 2\mathbb{P}_i - 3 \mathcal{K} + \frac{7}{4} \right] +\mathbb{P}_i \left(\mathbb{M}_i - 3\mathcal{K} + \frac{1}{4}\right) + \mathbb{M}_i \left(\frac{1}{4} -\mathcal{K}\right) \nonumber \\
&& - P_i\sum_{k\neq i }^{N} (P_k +  \mathbb{P}_k - 2\mathbb{M}_k) - \mathbb{P}_i\sum_{k\neq i }^{N}(P_k +  \mathbb{P}_k - 2\mathbb{M}_k) - \mathbb{M}_i\sum_{k\neq i }^{N}(P_k +  \mathbb{P}_k), \\
\mathcal{K}_{, \rho} &=&  2 \mathcal{K} \left( \frac{1}{4}-\mathcal{K} + \sum_{k= 1 }^N\mathbb{M}_k \right), \\
\mathbb{Q}_{i, \rho}&=& \mathbb{Q}_i \left[  2\mathcal{K} - 2 \sum_{k= 1 }^N\mathbb{M}_k-\frac{3}{2}\right]-  \sum_{i\neq k}^N\mathbb{J}_u^{(i,k)},
\end{eqnarray}
where, for notational reasons we have set
\begin{equation}
\mathbb{J}_u^{(i,k)}=\frac{j_u^{(i,k)}}{{\phi}^3},\quad \mathbb{J}_e^{(i,k)}= \frac{j_e^{(i,k)}}{{\phi}^3}.
\end{equation}
In the case of two fluids the equations above reduce to
\begin{eqnarray}
\hat{\phi}&=&-\frac{1}{2} {\phi}^2 - \frac{2}{3} ({\mu}_1 + {\mu}_2) - \mathcal{E} - \frac{1}{2} ({\Pi}_1  + {\Pi}_2),  \\
\hat{\mathcal{E}}&=&  \frac{1}{3} (\hat{\mu}_1 - \hat{\mu}_2) - \frac{1}{2} ( \hat{\Pi}_1 + \hat{\Pi}_2)-\frac{3}{2} {\phi} \left(\mathcal{E} + \frac{1}{2} {\Pi}_1 + \frac{1}{2} {\Pi}_2 \right),  \\ 
-\mathcal{A} \phi &+& \frac{1}{3}({\mu}_1 + 3 p_1) + \frac{1}{3}({\mu}_2 + 3 p_2)- \mathcal{E} + \frac{1}{2} {\Pi}_1 + \frac{1}{2} {\Pi}_2 = 0,\\
\hat{\mathcal{A}}&=&- \mathcal{A} \left( \mathcal{A} + \phi \right) + \frac{1}{2}({\mu}_1 + 3p_1) + \frac{1}{2}({\mu}_2 + 3p_2), \\
 \hat{p}_1 + \hat{\Pi}_1&=&-{\Pi}_1 \left(\frac{3}{2} \phi + \mathcal{A} \right) - \mathcal{A} \left( {\mu}_1 + p_1 \right) + j_e, \\
 \hat{p}_2 + \hat{\Pi}_2 &=&-{\Pi}_2 \left(\frac{3}{2} \phi + \mathcal{A} \right) - \mathcal{A} \left( {\mu}_2 + p_2 \right) - j_e ,\\
K&=&\frac{1}{3} ({\mu}_1 + {\mu}_2) - \mathcal{E} - \frac{1}{2} ({\Pi}_1 + {\Pi}_2) + \frac{1}{4} {\phi}^2 ,\\
\hat{Q}_1&=&-Q_1 \left(\phi + 2 \mathcal{A} \right) + j_u, \\
Q_2 &=& - Q_1, 
\end{eqnarray}
and in terms of $\rho$, we have
\begin{subequations}
\begin{align} 
\phi{\phi}_{,\rho}=&-\frac{1}{2} {\phi}^2 - \frac{2}{3} ({\mu}_1 + {\mu}_2) - \mathcal{E} - \frac{1}{2} ({\Pi}_1  + {\Pi}_2),  \\
{\mathcal{E}}_{, \rho }=&  \frac{1}{3} ({\mu}_1 - {\mu}_2) - \frac{1}{2} ({\Pi}_1 + {\Pi}_2)-\frac{3}{2}  \left(\mathcal{E} + \frac{1}{2} {\Pi}_1 + \frac{1}{2} {\Pi}_2 \right), \\
-\mathcal{A} \phi &+ \frac{1}{3}({\mu}_1 + 3 p_1) + \frac{1}{3}({\mu}_2 + 3 p_2) - \mathcal{E} + \frac{1}{2} {\Pi}_1 + \frac{1}{2} {\Pi}_2 = 0,\\
\phi{\mathcal{A}}_{, \rho }=&- \mathcal{A} \left( \mathcal{A} + \phi \right) + \frac{1}{2}({\mu}_1 + 3p_1) + \frac{1}{2}({\mu}_2 + 3p_2), \\
 \phi{p}_{1, \rho } + \phi{\Pi}_{1, \rho }=&-{\Pi}_1 \left(\frac{3}{2} \phi + \mathcal{A} \right) - \mathcal{A} \left( {\mu}_1 + p_1 \right) + j_e^{(1,2)}, \\
 \phi{p}_{2, \rho }+ \phi{\Pi}_{2, \rho }=&-{\Pi}_2 \left(\frac{3}{2} \phi + \mathcal{A} \right) - \mathcal{A} \left( {\mu}_2 + p_2 \right) - j_e^{(1,2)}, \\
K+ \mathcal{E}-\frac{1}{4} {\phi}^2 =&\frac{1}{3} ({\mu}_1 + {\mu}_2) - \frac{1}{2} ({\Pi}_1 + {\Pi}_2)  ,\\
\phi{Q}_{1, \rho}=&-Q_1 \left(\phi + 2 \mathcal{A} \right) + j_u^{(1,2)},  \\
Q_2 =& - Q_1.
\end{align}
\end{subequations}
The corresponding TOV equations are given in \eqref{tov2}.
\end{widetext}
%%%%%%%%%%%%%%%%%%%%%%%%%%%%%%%%%%%%%%%%%%%%%%%%%%%%%%%%%%%%%%%%%%%%%%%
\begin{widetext}
\section{Full solutions for Sec. \ref{RS}} \label{AARS}
%%%%%%%%%%%%%%%%%%%%%%%%%%%%%%%%%%%%%%%%%%%%%%%%%%%%%%%%%%%%%%%%%%%%%%%%
\subsection{Sec. \ref{FL2}: The case $\mathbb{P}_1 \neq 0$}
\begin{subequations}\label{f4a}
\begin{align}
p_{r1} &=\frac{1}{4 a-4 b r^2} +\frac{1}{3} e^{\alpha } c_1 \left[12 a^2+3 a (1-4 b) r^2 +2 b (2 b-1) r^4\right]^{\frac{5-4 b}{16 b-8}},\\
p_{\perp 1} &=\left\{-3 a^2 b [8 b (26 b-67)+141] r^4+9 a^3 [24 b (2 b-7) +19] r^2+576 a^4+4 a b^2 [40 b (2 b-5)+89] r^6 \right. \nonumber \\
&\left. -4 (5-4 b)^2 b^3 r^8\right\}\times \left\{192 \left(a-b r^2\right)^3 \left[12 a^2 +3 a (1-4 b) r^2+2 b (2 b-1) r^4\right]\right\}^{-1}+\frac{1}{3} e^{\alpha } c_1 \left[12 a^2 \right. \nonumber \\
&\left. +3 a (1-4 b) r^2+2 b (2 b-1) r^4\right]^{\frac{5-4 b}{16 b-8}},\\
p_{r2} &= \frac{3}{4 \left(a-b r^2\right)}-\frac{1}{3} e^{\alpha } c_1 \left[12 a^2+3 a (1-4 b) r^2 +2 b (2 b-1) r^4\right]^{\frac{5-4 b}{16 b-8}},\\
p_{\perp 2} &=\left\{(4 b r+r)^2 \left(3 a-2 b r^2\right) \left(9 a^2-7 a b r^2+2 b^2 r^4\right) -64 e^{\alpha } c_1 \left(a-b r^2\right)^3 \left[12 a^2+3 a (1-4 b) r^2 \right. \right. \nonumber \\
&\left. \left. +2 b (2 b-1) r^4\right]^{\frac{3}{8} \left(\frac{1}{2 b-1}+2\right)}-6 a^2 b [8 b (26 b-107) +205] r^4+54 a^3 [8 b (2 b-11)+9] r^2+1728 a^4 \right. \nonumber \\
&\left. +8 a b^2 [80 (b-4) b+131] r^6-8 b^3 [16 (b-4) b +37] r^8\right\} \times \left\{192 \left(a-b r^2\right)^3 \left[12 a^2+3 a (1-4 b) r^2 +2 b (2 b-1) r^4\right]\right\}^{-1}.
\end{align}
\end{subequations}
%%%%%%%%%%%%%%%%%%%%%%%%%%%%%%%%%%%%%%%%%%%%%%%%%%%%%%%%%%%%%%
\subsection{Sec. \ref{FL3}: The case $\mathbb{P}_1 = 0$}
\begin{subequations} \label{feos1a}
\begin{align}
p_{r1} = p_{\perp 1} &= \frac{1}{3} e^{\beta } c_1 \left[12 a^2+3 a (1-4 b) r^2 +2 b (2 b-1) r^4\right]^{\mathfrak{z}_3},\\
p_{r2} &= \frac{1}{a-b r^2}-\frac{1}{3} e^{\beta } c_1 \left[12 a^2+3 a (1-4 b) r^2 +2 b (2 b-1) r^4\right]^{\mathfrak{z}_3},\\
p_{\perp 2} &= \left\{r^2 (4 b r+r)^2 \left(3 a-2 b r^2\right) \left(9 a^2-7 a b r^2 +2 b^2 r^4\right)-48 e^{\beta } c_1 r^2 \left(a-b r^2\right)^3 \left[12 a^2 \right. \right. \nonumber \\
&\left. \left.+3 a (1-4 b) r^2+2 b (2 b-1) r^4\right]^{\frac{12 b+6 \gamma -3}{16 b-8}}-6 a^2 b [8 b (26 b-107)+205] r^6\right. \nonumber \\
&\left. +54 a^3 [8 b (2 b-11)+9] r^4+1728 a^4 r^2+8 a b^2 [80 (b-4) b+131] r^8\right. \nonumber \\
&\left.-8 b^3 [16 (b-4) b+37] r^{10}\right\} \times \left\{144 r^2 \left(a-b r^2\right)^3 \left[12 a^2+3 a (1-4 b) r^2+2 b (2 b-1) r^4\right]\right\}^{-1},
\end{align}
\end{subequations}
where 
\begin{equation}\label{beta}
\beta =  \frac{\sqrt{3} (4 b+1) (2 \gamma +1) \tan ^{-1}\left[\frac{a (3-12 b)+4 b (2 b-1) r^2}{a \sqrt{48 b^2-24 b-9}}\right]}{4 (2 b-1) \sqrt{(4 b-3) (4 b+1)}},
\end{equation}
and we have used \eqref{fz3} for the case $\mathbb{P}_1 = 0$. 
%%%%%%%%%%%%%%%%%%%%%%%%%%%%%%%%%%%%%%%%%%%%%%%%%%%%%%%%%%%%%%
\subsection{Sec. \ref{FL3}: The case $\mathbb{P}_1 \neq 0$}
%%%%%%%%%%%%%%%%%%%%%%%%%%%%%%%%%%%%%%%%%%%%%%%%%%%%%%%%%%%%%%
\begin{subequations}\label{feos2a}
\begin{align}
p_{r1} &= \frac{1}{3} e^{\beta } c_1 \left[12 a^2+3 a (1-4 b) r^2 +2 b (2 b-1) r^4\right]^{\mathfrak{z}_3},\\
p_{\perp 1} &= \frac{1}{12} e^{\beta } c_1 \left[12 a^2+3 a (1-4 b) r^2 +2 b (2 b-1) r^4\right]^{\frac{-20 b+6 \gamma +13}{16 b-8}} \left\{48 a^2  -3 a r^2 [4 b (\gamma +4)+\gamma -4]\right. \nonumber \\
&\left. +2 b r^4 [4 b (\gamma +2)+\gamma -4]\right\},\\
p_{r2} &= \frac{1}{a-b r^2}-\frac{1}{3} e^{\beta } c_1 \left[12 a^2+3 a (1-4 b) r^2 +2 b (2 b-1) r^4\right]^{\mathfrak{z}_3},\\
p_{\perp 2} &= \left\{-624 a^2 b^3 r^4+1608 a^2 b^2 r^4+432 a^3 b^2 r^2 -4 e^{\beta } c_1 \left(a-b r^2\right)^3 \left\{48 a^2-3 a r^2 [4 b (\gamma +4) \right. \right. \nonumber \\
&\left. \left.+\gamma -4]+2 b r^4 [4 b (\gamma +2)+\gamma -4]\right\} \left[12 a^2 +3 a (1-4 b) r^2+2 b (2 b-1) r^4\right]^{\mathfrak{z}_3}\right. \nonumber \\
&\left.-423 a^2 b r^4-1512 a^3 b r^2+171 a^3 r^2+576 a^4 +320 a b^4 r^6-800 a b^3 r^6+356 a b^2 r^6-64 b^5 r^8\right. \nonumber \\
&\left.+160 b^4 r^8-100 b^3 r^8\right\} \times \left\{48 \left(a-b r^2\right)^3 \left[12 a^2 +3 a (1-4 b) r^2+2 b (2 b-1) r^4\right]\right\}^{-1},
\end{align}
\end{subequations}
where we have used \eqref{fz3}.
%%%%%%%%%%%%%%%%%%%%%%%%%%%%%%%%%%%%%%%%%%%%%%%%%%%%%%%%%%%%%%
\section{Full solutions for Sec. \ref{SR}: Interior Schwarzschild-Tolman IV} \label{AASR}
%%%%%%%%%%%%%%%%%%%%%%%%%%%%%%%%%%%%%%%%%%%%%%%%%%%%%%%%%%%%%%%%%%%%%%%%%%%%%
\begin{subequations}\label{RTS0}
\begin{align}
P_1 =&  \left[c_1 e^{\rho } \left(A^2+2 e^{\rho }\right) \left(\frac{c_1+z_1}{c_1}\right)^{\frac{1}{2} (-a-1)} \left(1-\frac{z_1}{c_1}\right)^{\frac{a+1}{2}}\left(z_1^2-c_1^2\right)^{\frac{1}{2} (-a-1)}\right]\times \left[\left(A^2+e^{\rho }\right) \left(e^{\rho }-R^2\right)\right]^{-1}\\
P_2 =& \frac{1}{12} \left(\frac{2 e^{\rho } \left(A^4+A^2 \left(2 e^{\rho }+R^2\right)+2 e^{2 \rho }\right) \left(-c_1 z_1+2 \mu_1 e^{\rho }-3\right)}{\left(A^2+e^{\rho }\right) \left(A^2+2 e^{\rho }\right) \left(e^{\rho }-R^2\right) \left(-c_1 z_1+\mu_1 e^{\rho }-3\right)} \frac{R^2 \left(A^2+2 e^{\rho }\right)}{\left(A^2+e^{\rho }\right) \left(e^{\rho }-R^2\right)}+\frac{2 \mu_1^2 e^{2 \rho }}{z_1^2 (c_1+z_1)^2} \right. \nonumber \\
& \left. +\frac{2 \mu_1 e^{\rho } \left(c_1 \left(\mu_1 e^{\rho }-6\right)-6 z_1\right)}{\left(z_1^2\right)^{3/2} (c_1+z_1)^2}-\frac{2 \mu_1 e^{\rho }}{c_1 z_1+z_1^2}+1\right)-\frac{c_1 e^{\rho } \left(A^2+2 e^{\rho }\right) \left(\frac{c_1+z_1}{c_1}\right)^{\frac{1}{2} (-a-1)} \left(1-\frac{z_1}{c_1}\right)^{\frac{a+1}{2}} \left(z_1^2-c_1^2\right)^{\frac{1}{2} (-a-1)}}{\left(A^2+e^{\rho }\right) \left(e^{\rho }-R^2\right)}\\
\mathbb{P}_2 =& \left[e^{2 \rho } \left(-A^4 \mu_1 (c_1+z_1) \left(\mu_1 R^2-3\right)+A^2 \left(c_1^2 \left(z_1^2\right)^{3/2}-2 \mu_1 e^{\rho } \left(2 \mu_1 R^2 (c_1+z_1)+3 c_1\right)+3 \mu_1 R^2 (c_1+z_1) \right. \right. \right. \nonumber \\
& \left. \left. \left. +\mu_1^2 e^{2 \rho } (3 c_1+2 z_1)+9 (2 c_1+z_1)\right)+2 \left(c_1^2 R^2 \left(z_1^2\right)^{3/2}-6 \mu_1 e^{\rho } R^2 (2 c_1+z_1)+\mu_1 e^{2 \rho } \left(c_1 \mu_1 R^2+3 (c_1+z_1)\right) \right. \right. \right. \nonumber \\
& \left. \left. \left. +9 R^2 (2 c_1+z_1)\right)\right)\right]\times \left[6 \left(z_1^2\right)^{3/2} \left(A^2+e^{\rho }\right) \left(A^2+2 e^{\rho }\right) (c_1+z_1)^2 \left(e^{\rho }-R^2\right)\right]^{-1},
\end{align}
\end{subequations}
where 
\begin{equation}\label{z1}
z_1 = \sqrt{3-\mu_1 e^{\rho }}
\end{equation}

\begin{subequations}\label{ST0}
\begin{align}
p_{r1} =p_{\perp 1} &= -\frac{4 \mathfrak{c}_1}{R^2} \left(\frac{c_1+\mathfrak{z}_4}{c_1}\right)^{-\mathfrak{z}_5} \left(1-\frac{\mathfrak{z}_4}{c_1}\right)^{\mathfrak{z}_5} \left(\mathfrak{z}_4^2-c_1^2\right)^{-\mathfrak{z}_5},\\
p_{r2} &=-\left\{-2 r^4 \left[-A^2 \left\{c_1^2 \mathfrak{z}_4^3+c_1 \left[\mu_1 r^2 \left(3 \mu_1 r^2 -4 \mu_1 R^2-6\right)+3 \mu_1 R^2+18\right]\right. \right. \right. \nonumber \\
& \left. \left. \left. +\mathfrak{z}_4 \left\{\mu_1 \left[2 \mu_1 r^4+R^2 \left(3-4 \mu_1 r^2\right)\right]+9\right\}\right\}+A^4 \mu_1 (c_1+\mathfrak{z}_4) \left(\mu_1 R^2-3\right)-2 c_1 \mu_1^2 r^4 R^2 \right. \right. \nonumber \\
& \left. \left. -6 \mu_1 r^2 \left[c_1 \left(r^2-4 R^2\right)+\mathfrak{z}_4 \left(r^2-2 R^2\right)\right] -2 R^2 \left[c_1 \left(c_1 \mathfrak{z}_4^3+18\right)+9 \mathfrak{z}_4\right]\right]\right. \nonumber \\
& \left. -12 \mathfrak{c}_1 r^2 \mathfrak{z}_4^3 \left(A^2+2 r^2\right)^2 (c_1+\mathfrak{z}_4)^2 \left(\frac{c_1+\mathfrak{z}_4}{c_1}\right)^{-\mathfrak{z}_5} \left(1-\frac{\mathfrak{z}_4}{c_1}\right)^{\mathfrak{z}_5} \left(\mathfrak{z}_4^2-c_1^2\right)^{-\mathfrak{z}_5}+2 r^2 \mathfrak{z}_4^2 (c_1+\mathfrak{z}_4) \right. \nonumber \\
& \left. \left[A^2 \left(2 r^2+R^2\right)+A^4+2 r^4\right] \left(c_1 \mathfrak{z}_4-2 \mu_1 r^2+3\right)+R^2 \mathfrak{z}_4^3 \left(A^2+2 r^2\right)^2 (c_1+\mathfrak{z}_4)^2\right. \nonumber \\
& \left. -2 \mu_1 r^2 \mathfrak{z}_4^2 \left(A^2+r^2\right) \left(A^2+2 r^2\right) (c_1+\mathfrak{z}_4) (r^2-R^2)+\mathfrak{z}_4^3 \left(A^2+r^2\right) \left(A^2+2 r^2\right) (c_1+\mathfrak{z}_4)^2 (r^2-R^2)\right. \nonumber \\
& \left. +2 \mu_1 r^2 \left(A^2+r^2\right) \left(A^2+2 r^2\right) (r^2-R^2)  \left(c_1 \mu_1 r^2-6 (c_1+\mathfrak{z}_4)\right)+2 \mu_1^2 r^4 \mathfrak{z}_4 \left(A^2+r^2\right) \right. \nonumber \\
& \left. \left(A^2+2 r^2\right) (r^2-R^2)\right\} \times \left\{3 r^2 R^2 \mathfrak{z}_4^3 \left(A^2+2 r^2\right)^2 (c_1+\mathfrak{z}_4)^2\right\}^{-1},\\
p_{\perp 2} &= -\left\{r^4 \left[-A^2 \left\{c_1^2 \mathfrak{z}_4^3+c_1 \left[\mu_1 r^2 \left(3 \mu_1 r^2 -4 \mu_1 R^2-6\right)+3 \mu_1 R^2+18\right]\right. \right. \right. \nonumber \\
& \left. \left. \left. +\mathfrak{z}_4 \left\{\mu_1 \left[2 \mu_1 r^4+R^2 \left(3-4 \mu_1 r^2\right)\right]+9\right\}\right\}+A^4 \mu_1 (c_1+\mathfrak{z}_4) \left(\mu_1 R^2-3\right)-2 c_1 \mu_1^2 r^4 R^2\right. \right. \nonumber \\
& \left. \left.-6 \mu_1 r^2 \left[c_1 \left(r^2-4 R^2\right)+\mathfrak{z}_4 \left(r^2-2 R^2\right)\right]-2 R^2 \left[c_1 \left(c_1 \mathfrak{z}_4^3+18\right)+9 \mathfrak{z}_4\right]\right]\right. \nonumber \\
& \left. -12 \mathfrak{c}_1 r^2 \mathfrak{z}_4^3 \left(A^2+2 r^2\right)^2 (c_1+\mathfrak{z}_4)^2 \left(\frac{c_1+\mathfrak{z}_4}{c_1}\right)^{-\mathfrak{z}_5} \left(1-\frac{\mathfrak{z}_4}{c_1}\right)^{\mathfrak{z}_5} \left(\mathfrak{z}_4^2-c_1^2\right)^{-\mathfrak{z}_5}+2 r^2 \mathfrak{z}_4^2 (c_1+\mathfrak{z}_4) \right. \nonumber \\
& \left. \left[A^2 \left(2 r^2+R^2\right)+A^4+2 r^4\right] \left(c_1 \mathfrak{z}_4-2 \mu_1 r^2+3\right)+R^2 \mathfrak{z}_4^3 \left(A^2+2 r^2\right)^2 (c_1+\mathfrak{z}_4)^2 \right. \nonumber \\
& \left. -2 \mu_1 r^2 \mathfrak{z}_4^2 \left(A^2+r^2\right) \left(A^2+2 r^2\right) (c_1+\mathfrak{z}_4) (r^2-R^2)+\mathfrak{z}_4^3 \left(A^2+r^2\right) \left(A^2+2 r^2\right) (c_1+\mathfrak{z}_4)^2 (r^2-R^2)\right. \nonumber \\
& \left. +2 \mu_1 r^2 \left(A^2+r^2\right) \left(A^2+2 r^2\right) (r^2-R^2) \left(c_1 \mu_1 r^2-6 (c_1+\mathfrak{z}_4)\right)+2 \mu_1^2 r^4 \mathfrak{z}_4 \left(A^2+r^2\right) \left(A^2+2 r^2\right)\right. \nonumber \\
& \left.  (r^2-R^2)\right\} \times \left\{3 r^2 R^2 \mathfrak{z}_4^3 \left(A^2+2 r^2\right)^2 (c_1+\mathfrak{z}_4)^2\right\}^{-1}, 
\end{align}
\end{subequations}
where
\begin{subequations}\label{z12r}
\begin{align}
\mathfrak{z}_4 = &\sqrt{3-r^2 \mu_1},\\
\mathfrak{z}_5 = & \frac{a+1}{2}.
\end{align}
\end{subequations}
The energy densities of both fluids are
\begin{subequations}\label{f10ed}
\begin{align}
\mu_{1} =& -\frac{4 a \mathfrak{c}_1}{R^2} \left(-c_1^2-\mu_1 r^2+3\right)^{\frac{1}{2} (-a-1)} \left(\frac{c_1+\mathfrak{z}_4}{c_1}\right)^{\frac{1}{2} (-a-1)} \left(1-\frac{\mathfrak{z}_4}{c_1}\right)^{\frac{a+1}{2}},\\
\mu_{2} =& -\left[-4 a \mathfrak{c}_1 \left(A^2+2 r^2\right)^2 \left(-c_1^2-\mu_1 r^2+3\right)^{\frac{1}{2} (-a-1)} \left(\frac{c_1+\mathfrak{z}_4}{c_1}\right)^{\frac{1}{2} (-a-1)} \left(1-\frac{\mathfrak{z}_4}{c_1}\right)^{\frac{a+1}{2}} \right. \nonumber\\
& \left. -r^2 \left(7 A^2+2 R^2\right)-3 A^2 \left(A^2+R^2\right)-6 r^4 \right] \times \left\{R^2 \left(A^2+2 r^2\right)^2\right\}^{-1},
\end{align}
\end{subequations}
where we have used \eqref{z12r}.
\end{widetext}

\end{document}